\documentclass{aa} 

\usepackage{psfig}

\begin{document}
\thesaurus{08            
              (08.06.2;  
               08.09.2 Elias~29;  
               09.04.1;  
               09.13.2;  
               09.01.1;  
               13.09.4)} 
         
\title{Infrared observations of hot gas and cold ice toward the low 
       mass protostar \object{Elias~29}\thanks{Based on
  observations with ISO, an ESA project with instruments
  funded by ESA Member States (especially the PI countries:
  France, Germany, the Netherlands and the United Kingdom)
  and with the participation of ISAS and NASA.}  }

\author{A.C.A. Boogert\inst{1,2,3} 
        \and A.G.G.M. Tielens\inst{1,2}
        \and C. Ceccarelli\inst{4}
        \and A.M.S. Boonman\inst{5} 
        \and E.F. van Dishoeck\inst{5} 
        \and J.V. Keane\inst{1}
        \and D.C.B. Whittet\inst{6}
        \and Th. de Graauw\inst{2}}
\offprints{\\ A.C.A. Boogert (boogert@submm.caltech.edu)}

\institute{Kapteyn Astronomical Institute, P.O. Box 800, 
                9700  AV Groningen, the Netherlands
           \and
           SRON, P.O. Box 800, 9700 AV Groningen, the Netherlands
           \and
           Present address: California Institute of Technology,
              Downs Laboratory of Physics 320-47, Pasadena, CA 91125,
              USA
           \and
           Laboratoire d'Astrophysique de
              l'Observatoire de Grenoble, B.P. 53X, 38041 Grenoble
              Cedex, France
           \and
           Leiden Observatory, P. O. Box 9513, 2300 RA Leiden, 
                                        the Netherlands
           \and
           Department of Physics, Applied Physics \& Astronomy,
                Rensselaer Polytechnic Institute, Troy, NY 12180, USA}

\date{Received 28 February 2000/ Accepted April 2000}

\maketitle\markboth{A.C.A. Boogert et al.: The low mass protostar
Elias 29}{}

\begin{abstract}
 
  We have obtained the full 1-200~${\rm \mu m}$ spectrum of the low
  luminosity (36~$\rm L_{\odot}$) Class~I protostar \object{Elias~29}
  in the $\rho$~Ophiuchi molecular cloud. It provides a unique
  opportunity to study the origin and evolution of interstellar ice
  and the interrelationship of interstellar ice and hot core gases
  around low mass protostars. We see abundant hot CO and ${\rm H_2O}$
  gas, as well as the absorption bands of CO, ${\rm CO_2}$, ${\rm
    H_2O}$ and ``6.85~${\rm \mu m}$'' ices.  We compare the abundances
  and physical conditions of the gas and ices toward \object{Elias~29}
  with the conditions around several well studied luminous, high mass
  protostars. The high gas temperature and gas/solid ratios resemble
  those of relatively evolved high mass objects (e.g. GL~2591).
  However, none of the ice band profiles shows evidence for
  significant thermal processing, and in this respect
  \object{Elias~29} resembles the least evolved luminous protostars,
  such as \object{NGC~7538~:~IRS9}.  Thus we conclude that the heating
  of the envelope of the low mass object \object{Elias~29} is
  qualitatively different from that of high mass protostars.  This is
  possibly related to a different density gradient of the envelope or
  shielding of the ices in a circumstellar disk. This result is
  important for our understanding of the evolution of interstellar
  ices, and their relation to cometary ices.


\end{abstract}

\section{Introduction}~\label{se29:intro}

The general picture of low mass star formation has been formed since
the 1980's with the availability of infrared and millimeter wavelength
broad band photometry from the ground, and with the IRAS satellite and
KAO observatory (e.g., Lada \& Wilking \cite{lad84}; Adams et al.
\cite{ada87}; Hillenbrand et al. \cite{hil92}; Andr\'e et al.
\cite{and93}).  A classification scheme was made, where the continuum
emission of Class~0 and~I objects peaks in the submillimeter and
far-infrared. These objects are still deeply embedded in their
accreting envelopes.  In the Class~II phase, the wind of the protostar
has cleared its surrounding environment, such that it becomes
optically visible, and shows \ion{H}{i} emission lines.  The continuum
emission of these objects peaks in the near-infrared, but there is
still significant excess emission above the stellar continuum. They
are believed to be surrounded by optically thick dusty disks. Finally,
little dust emission remains for Class~III objects, when the disk is
optically thin, and planetary companions may have been formed.

Our knowledge of the physical and chemical state and evolution of the
material surrounding protostars, has progressed with the availability
of medium and high resolution spectroscopic instrumentation at near
and mid-infrared wavelengths ($\sim2-20$~${\rm \mu m}$).  The progress
made, is best illustrated by the observations of high mass protostars,
which are bright and easy to observe.  The (ro-)vibrational bands of
various molecules (CO, ${\rm H_2O}$, ${\rm CH_3OH}$, silicates) were
observed from the ground, revealing profile variations of the
3.07~${\rm \mu m}$, and 4.67~${\rm \mu m}$ ${\rm H_2O}$ and CO ice
bands (e.g. Smith et al. \cite{smi89}; Tielens et al. \cite{tie91};
Chiar et al. \cite{chi98}). This was interpreted as evaporation of the
volatile CO ice, and crystallization of ${\rm H_2O}$ ice in the
molecular envelopes. This heating effect is strengthened by the
detection of hot gas in 4.6~${\rm \mu m}$ CO observations (Mitchell et
al. \cite{mit91}). With the launch of the {\it Infrared Space
  Observatory} in 1995 (ISO; Kessler et al. \cite{kes96}), it became
possible to observe all other molecular bands in the infrared (Whittet
et al. \cite{whi96}). It was shown that high mass protostellar
evolution can be traced in the gas-to-solid abundance ratios (van
Dishoeck et al.  \cite{dis96}; van Dishoeck \& Blake \cite{disb98}),
and the profiles of the ice bands, in particular solid ${\rm CO_2}$
(Gerakines et al. \cite{ger99}; Boogert et al. \cite{boo00}).  Thus,
there is overwhelming evidence that thermal processing, i.e.
evaporation and crystallization of ices in and around hot molecular
cores, plays an important role in the evolution of high mass molecular
envelopes.

The composition and evolution of the molecular material around low
mass protostars are not as well studied.  It seems unlikely that the
molecular material evolves similar to that around high mass
protostars.  Low mass protostars evolve much slower, release less
radiative energy, drive less energetic winds, and form disks.  It is
not established whether low mass objects possess hot cores as well,
and whether the ices survive the process of star formation.  If (some
of) the ices survive, are they included into comets, and if so, are
the ice structure and composition still the same compared to
interstellar ices?  How important are energetic processes, such as
cosmic ray bombardment, in altering the ice composition on the long
time scale of the formation of low mass stars?

To investigate the influence of low mass protostars on their molecular
envelope, we make an infrared spectroscopic study \object{Elias~29},
also called WL~15 and YLW~7 (Elias \cite{eli78}, Wilking et al.
\cite{wil83}, Young et al.  \cite{you86}). On a large scale,
\object{Elias~29} lies in core E, which is in the south-east corner of
the 1$\times$2~pc extended compact CO ridge L~1688 (Loren et al.
\cite{lor90}) in the densest part of the $\rho$~Ophiuchi cloud, at a
distance of $\sim$160~pc from the earth (Wilking \& Lada \cite{wil83};
Whittet \cite{whi74}).  It is the reddest object found in the
near-infrared survey of this cloud by Elias (\cite{eli78}), without a
counterpart at optical wavelengths.  For our observations, we used
Elias' coordinates (J2000):

\begin{center}
  $\rm \alpha~=~16^h27^m09^s.3$ \hspace{20pt} $\rm
  \delta~=~-24^o37'21''$.
\end{center}

The overall spectrum of \object{Elias~29} is typical for a heavily
embedded Class~I source, probably in a late accretion phase (Wilking
et al. \cite{wil89}; Andr\'e \& Montmerle \cite{and94}; Greene \& Lada
\cite{gre96}; Saraceno et al. \cite{sar96}). The embedded nature is
also revealed by its high extinction, and by the cold compact envelope
observed at millimeter wavelengths (Andr\'e \& Montmerle \cite{and94};
Motte et al. \cite{mot98}).  \object{Elias~29} is associated with a
molecular outflow (Bontemps et al. \cite{bon96}; Sekimoto et al.
\cite{sek97}). With a bolometric luminosity of $\sim 36~L_{\odot}$
(Chen et al. \cite{che95}), \object{Elias~29} is the most luminous
protostar in the $\rho$~Oph cloud, which makes this source very
suitable for spectroscopic studies.  The relatively high luminosity,
and high bolometric temperature ($T_{\rm bol}\sim 410$~K) imply an age
in the range 0.5--4 10$^5$~yr (Chen et al.  \cite{che95}). In the
pre-main-sequence evolutionary tracks of Palla \& Stahler
(\cite{pal93}), this corresponds to a star with end mass 3.0--3.5
$M_{\odot}$. \object{Elias~29} might thus be a precursor Herbig AeBe
star. This classification is however uncertain.  For example, it has
been argued from the SED, and the absence of mid-infrared emission
features, that \object{Elias~29} is a 1 $M_{\odot}$ protostar with a
large accretion luminosity, and a spectral type of K3-4 at the birth
line (Greene \& Lada \cite{gre00}).

This Paper is structured as follows. Technical details on the ISO
infrared observations are given in Sect.~\ref{se29:obs}.  All the
observed emission and absorption features are discussed in detail in
Sect.~\ref{se29:res}.  Section~\ref{se29:sed} gives a description of
the continuum shape, and a comparison to other lines of sight. The ice
composition and thermal history, and the silicate band depth with
inferred extinction and column densities toward \object{Elias~29} are
discussed in Sect.~\ref{se29:ice}. Then, numerous lines of gaseous CO
and ${\rm H_2O}$ are detected, and modeled to derive gas temperatures
and column densities (Sect.~\ref{se29:gas}). The molecular abundances
and gas-to-solid ratios of \object{Elias~29} are compared to a sample
of sight-lines, ranging from dark cloud cores to evolved protostars. A
comparison with high mass protostars is made (Sect.~\ref{se29:abun}).
Section~\ref{se29:geom} discusses the origin of the wealth of observed
emission and absorption features and puts them in a geometrical
picture, where we review the evidence for an extended envelope and an
accretion disk. We conclude in Sect.~\ref{se29:summary} with a summary
and suggestions for future observations.

\section{Observations}~\label{se29:obs}

\subsection{The 2.3--45~$\bf \mu$m spectrum}~\label{se29:obssws}

A low resolution ($R=\lambda/\Delta \lambda=400$), full 2.3--45~${\rm
  \mu m}$ spectrum of \object{Elias~29} was obtained with the ISO {\it
  Short Wavelength Spectrometer} (ISO--SWS; de Graauw et al.
\cite{gra96}) during revolution 267 (August 10 1996). The ISO--SWS
pipeline and calibration files, available in July 1998 at SRON
Groningen, were applied. The spectrum is generally of good quality,
with well-matching up and down scans, and no serious dark current
problems, except for band 2C (7--12~${\rm \mu m}$).  Here, we found
that the up and down scans deviate over the silicate band. One scan
showed good agreement with a ground-based spectrum of Hanner et al.
(\cite{han95}), and we used this to correct the deviating scan.
Standard after-pipeline steps were applied, such as low order
flat-fielding, sigma clipping and re-binning (see also Boogert et al.
\cite{boo98}).  The twelve sub-spectra in the 2-45~${\rm \mu m}$ range
match fairly well at the overlap regions. Small correction factors
($<$15\%) were applied to correct for the band jumps.

At selected wavelength ranges (3--3.6, 4--9, and 19.5--28~${\rm \mu
  m}$), we also obtained high resolution ($R=1500$) ISO--SWS grating
spectra, in revolution 292 (September 04 1996). These were reduced
similarly to the low resolution spectrum. We found that the overall
shape of the spectrum near 4--5~${\rm \mu m}$ is quite badly affected
by detector memory effects, presumably due to the occurrence of scan
breaks (de Graauw et al.  \cite{gra96}). We corrected for this, by
applying a wavelength-dependent shift to match the low resolution
spectrum.  This does not affect our conclusions, since the high
resolution spectrum was only used to study narrow features. Also, near
6.9~${\rm \mu m}$ the scans deviate significantly because of memory
effects. This problem is reflected in the large error bars given in
this paper, as they were derived from the difference between the
average up and down scans.

\subsection{The 45--190~$\bf \mu$m spectrum}~\label{se29:obslws}   

\object{Elias~29} was observed during Revolution 484 (March 14 1997)
with the {\it ISO Long Wavelength Spectrometer} (ISO--LWS; Clegg et
al. \cite{cle96}).  We obtained 15 scans covering the range from
43~${\rm \mu m}$ to 197~${\rm \mu m}$ in the low resolution mode
($R\sim$200) for a total of 2611 sec of integration time. The data
were reduced using the Off-Line-Processing package (OLP) version~7 and
the ISO-Spectral-Analysis-Package (ISAP) version~1.3.  The spectra
were flux calibrated using Uranus (Swinyard et al.  \cite{swi96}). We
find that at the ISO--SWS/LWS overlap region near 45~${\rm \mu m}$,
LWS has 35\% higher flux than SWS.  This difference is only slightly
larger than the absolute calibration uncertainties of the two
instruments, and thus it is doubtful that this can be ascribed to the
presence of extended emission in the larger aperture of ISO--LWS
($\sim 80''$ versus $\sim 25''$).  We therefore decided to multiply
the LWS spectrum down with this factor.

\section{Results}~\label{se29:res}

\subsection{The spectral energy distribution (SED)}~\label{se29:sed}

\begin{figure*}[t!]
\begin{picture}(220,300)(0,0)
\psfig{figure=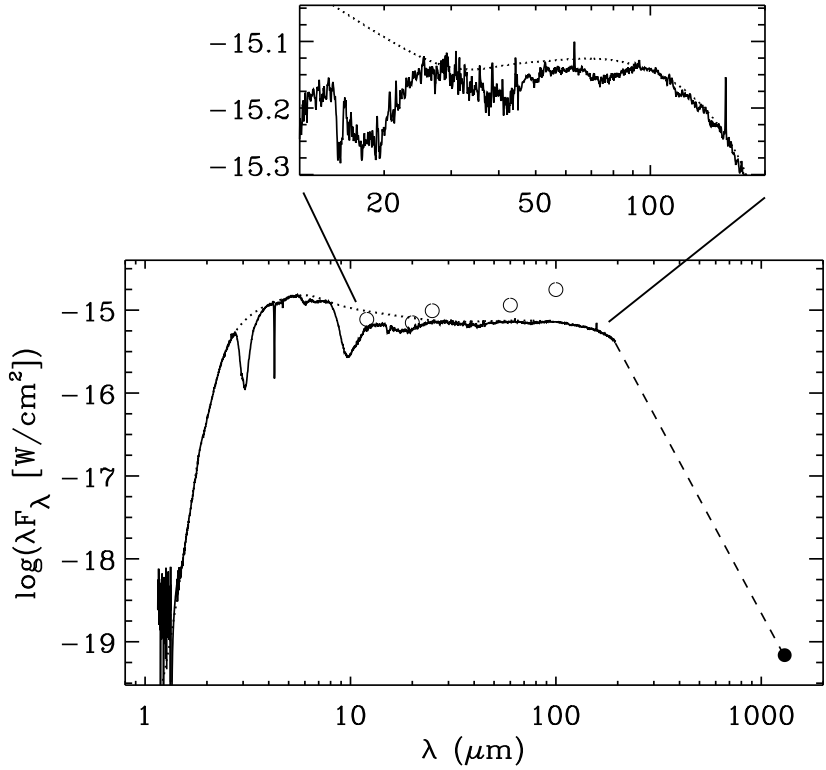,width=330pt,angle=90}
\end{picture}
\hfill \parbox[b]{160pt}{\caption{Spectral energy distribution of
    \object{Elias~29}, consisting of ground-based observations
    ($\lambda <$2.4~${\rm \mu m}$; Greene \& Lada \cite{gre96}), an
    ISO-SWS spectrum ($\lambda =$2.4--45~${\rm \mu m}$), and an
    ISO--LWS spectrum ($\lambda =$45--195~${\rm \mu m}$). The data
    point at 1300~${\rm \mu m}$ is taken from Andr\'e \& Montmerle
    (\cite{and94}), which we have connected with a dashed straight
    line to the ISO--LWS spectrum, to guide the eye. The dotted line
    is the adopted continuum, as determined by blackbody fits and by
    hand. The open circles are ground-based and IRAS observations (see
    text). The top inset shows a magnification of the 10-200~${\rm \mu
      m}$ region.}~\label{fe29:sed}}
\end{figure*}

\begin{figure*}[t!]
\begin{picture}(220,250)(0,0)
\psfig{figure=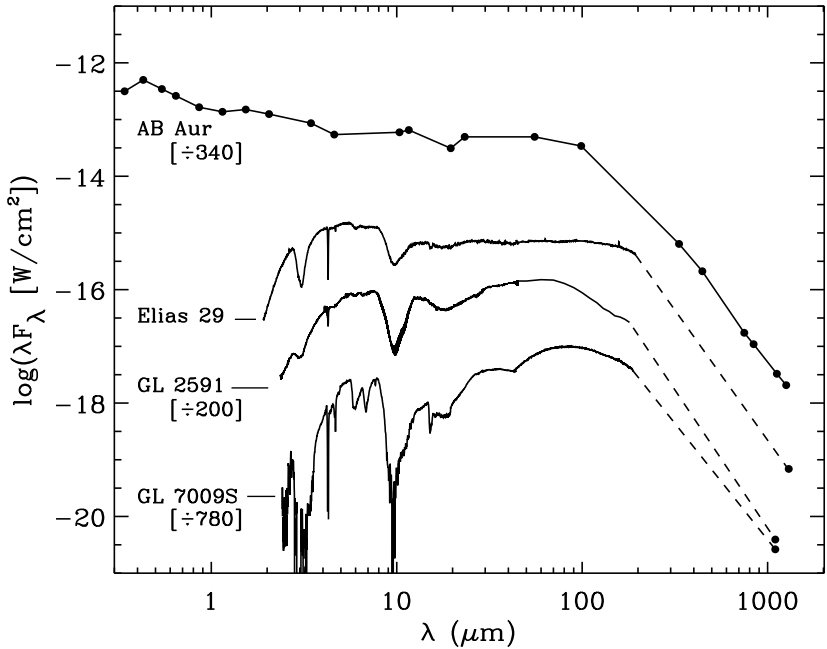,width=330pt,angle=90}
\end{picture}
\hfill \parbox[b]{160pt}{\caption[]{Spectral energy distribution of
    \object{Elias~29} compared to the high mass protostars GL~7009S
    (Dartois et al. \cite{dar98a}), and GL~2591 (van der Tak et al.
    \cite{tak99}). The spectrum of the Herbig Ae star AB~Aur is a
    compilation of continuum observations taken from Mannings
    (\cite{man94}), and is further discussed in Sect.~\ref{se29:geom}.
    The flux scale of each spectrum has been divided by the values
    given in brackets.}~\label{fe29:sedysos}}
\end{figure*}

\object{Elias~29} is only visible at wavelengths larger than
$\sim$1.5~${\rm \mu m}$ (Greene \& Lada \cite{gre96}; Elias
\cite{eli78}). Our ISO observations show that the continuum emission
rises steeply between 2--3~${\rm \mu m}$, reaches a maximum of
$\lambda F_{\lambda}$=15$\times10^{-16}$ W ${\rm cm^{-2}}$ at $\lambda
\sim$5~${\rm \mu m}$, and is remarkably flat with $\lambda
F_{\lambda}$$\sim 8\times10^{-16}$~W~${\rm cm^{-2}}$ between 20 and
100~${\rm \mu m}$ (Fig.~\ref{fe29:sed}).  The emission has dropped to
$\lambda F_{\lambda}$$\sim 4\times10^{-16}$~W~${\rm cm^{-2}}$ at
200~${\rm \mu m}$, and by four orders of magnitude at 1300~${\rm \mu
  m}$.  Our near-infrared spectral continuum fluxes are in excellent
agreement with broad band fluxes from ground-based observations (Elias
\cite{eli78}). Also the ground-based small beam 10 and 20~${\rm \mu
  m}$ observations, as well as the large beam 12 and 25~${\rm \mu m}$
IRAS fluxes, match the ISO--SWS observation well, thus indicating that
at these wavelengths the emission is well confined within a region of
8$''$ in diameter (Fig.~\ref{fe29:sed}; Lada \& Wilking \cite{lad84};
Young et al. \cite{you86}).  The reasonable match of the ISO--SWS and
LWS spectra (Sect.~2.2) indicates that also at 45~${\rm \mu m}$ the
emission is not very extended ($<25''$).  At 100~${\rm \mu m}$,
however, some large scale emission may be present, since the IRAS
flux, observed in a 5.5 times larger aperture, is a factor of 2 larger
compared to the ISO measurement.

The observed SED of \object{Elias~29} is different from that of
massive protostars such as GL~2591, and GL~7009S, which peak in the
far-infrared (Fig.~\ref{fe29:sedysos}). It has been proposed that the
shape of SEDs is independent of the luminosity of the central object,
and is rather determined by the total dust column density along the
line of sight (Ivezic \& Elitzur \cite{ive97}).  In the ``standard
model'' of Ivezic \& Elitzur, GL~2591 and GL~7009S would have a column
density corresponding to an $A_{\rm V}$ of several hundred. Elias 29
must have an $A_{\rm V}<100$, because it does not peak in the far
infrared.  However, the flatness of the SED up to 100~${\rm \mu m}$ is
not reproduced in these models.  A lower column density alone thus
cannot explain the differences between the SED of \object{Elias~29}
and massive protostars.  Other factors, such as a different density
gradient, and the presence of a circumstellar disk are probably
important.  We will discuss the structure of \object{Elias~29}, in
relation to the detected gas and ice absorption features, in
Sect.~\ref{se29:geom}.

\subsection{Ice and dust absorption bands}~\label{se29:ice}

\begin{figure*}[!t]
\begin{picture}(230,230)(-70,0)
\psfig{figure=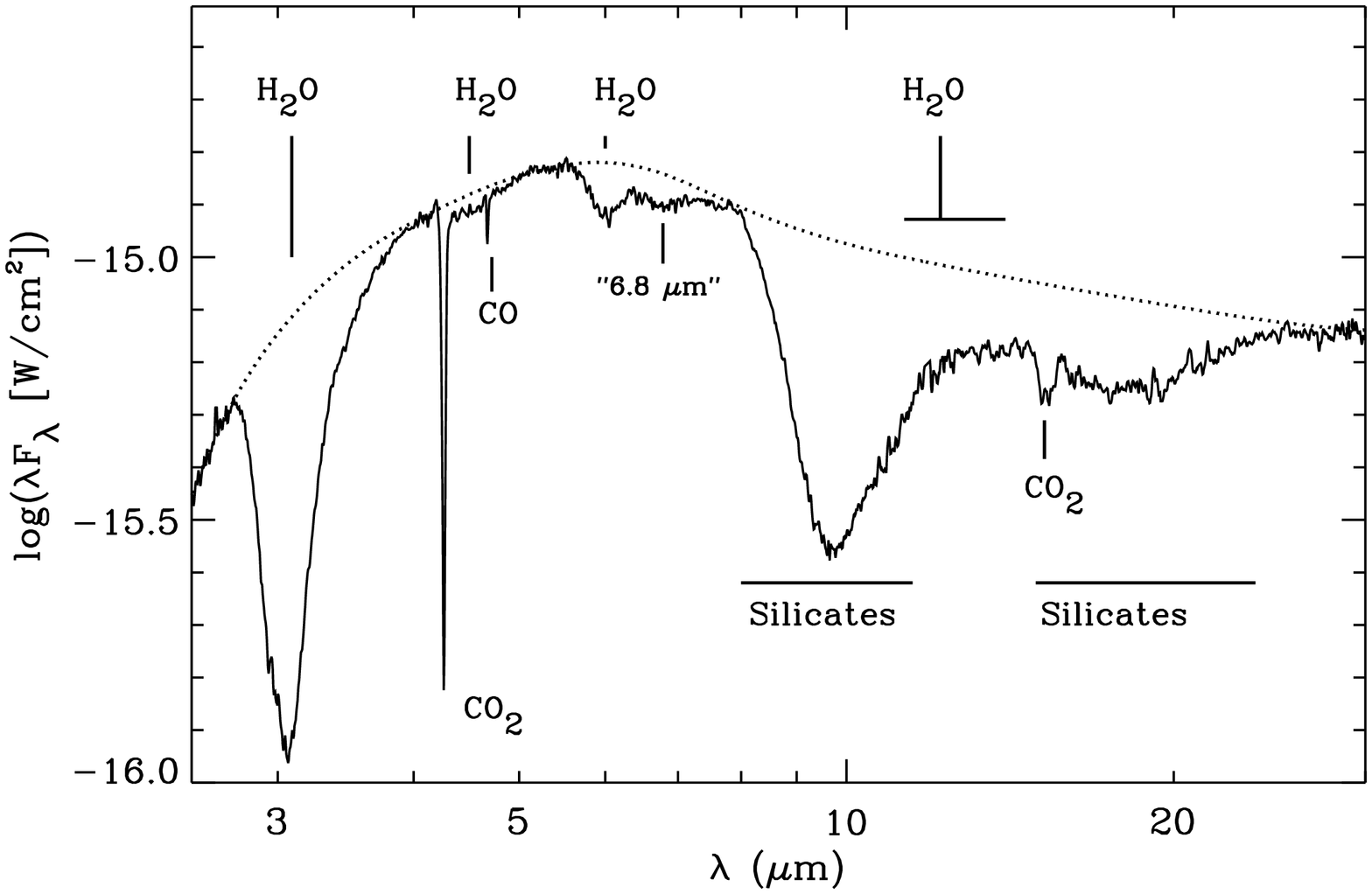,width=350pt,angle=90}
\end{picture}
\caption{Low resolution ($R=400$)
  mid-infrared ISO--SWS spectrum of \object{Elias~29} with a smooth,
  global continuum (dotted line), rather arbitrarily determined by
  hand and blackbody fits. The vibrational absorption bands of various
  molecules are indicated.}
\label{fe29:mir}
\end{figure*}

Numerous absorption bands of ices and silicates are present in the
infrared spectrum of \object{Elias~29} (Fig.~\ref{fe29:mir}), such
that hardly any 'clean' continuum emission is left.  We identify each
band, derive column densities, and, when possible, determine the ice
mantle composition and thermal history. The full spectrum also allows
a determination of upper limits of abundances for undetected, though
astrophysically relevant molecules. The column densities for the
species discussed below are summarized in Table~\ref{te29:colden}.

\begin{table}[!b]
\caption{Column densities of ices toward \object{Elias~29}. Non-detections are
indicated with 3$\sigma$ upper limits.}~\label{te29:colden}
\centering
\begin{tabular}{ll}
\noalign{\smallskip} 
\hline
\noalign{\smallskip} 
molecule         & $N~[10^{17}$~${\rm cm^{-2}}$] \\
\noalign{\smallskip} 
\hline
\noalign{\smallskip} 
${\rm H_2O}$     & 34 (6)        \\
${\rm ^{12}CO_2}$& 6.7 (0.5)     \\
${\rm ^{13}CO_2}$& 0.083 (0.005) \\
CO               & 1.7 (0.3)     \\
CH$_4$           & $< 0.5$       \\
NH$_3$           & $< 3.5$       \\
CH$_3$OH         & $< 1.5$       \\
H$_2$CO          & $< 0.6$       \\
HCOOH            & $< 0.3$       \\
OCS              & $< 0.015$     \\
XCN              & $< 0.067$     \\
\noalign{\smallskip} 
\hline
\end{tabular}
\end{table}

\subsubsection{H$_2$O ice}

\begin{figure}
\begin{picture}(220,197)(0,0)
\psfig{figure=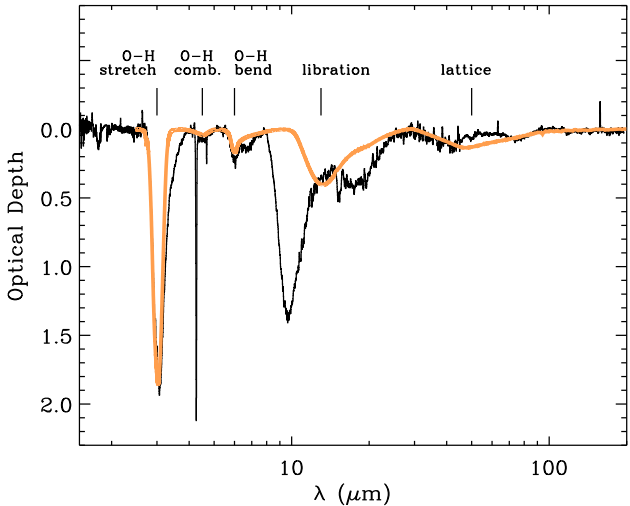,width=240pt,angle=90}
\end{picture}
\caption{Optical depth spectrum of \object{Elias~29}, assuming the continuum
  indicated in Figs.~\ref{fe29:sed} and~\ref{fe29:mir}. The light,
  thick line is a laboratory spectrum of ${\rm H_2O}$ ice at $T=$10~K
  (Hudgins et al. \cite{hud93}). All five ${\rm H_2O}$ ice vibration
  bands can be discerned.}~\label{fe29:waterfit}
\end{figure}

The infrared spectrum of \object{Elias~29} shows all the vibration
modes of ${\rm H_2O}$ ice in absorption (Fig.~\ref{fe29:waterfit}). We
see the O--H stretching mode at 3.0~${\rm \mu m}$ (``$\nu _1$, $\nu
_3$'' in spectroscopic notation), the O--H bending mode at 6.0~${\rm
  \mu m}$ (``$\nu _2$''), the libration or hindered rotation mode at
$\sim$12~${\rm \mu m}$ (``$\nu _{\rm L}$''), the combination mode at
4.5~${\rm \mu m}$ (``$3\nu _{\rm L}$'' or ``$\nu _2+\nu _{\rm L}$''),
and perhaps the lattice mode at $\sim$45~${\rm \mu m}$.

\begin{figure}
\begin{picture}(0,240)(0,0)
\psfig{figure=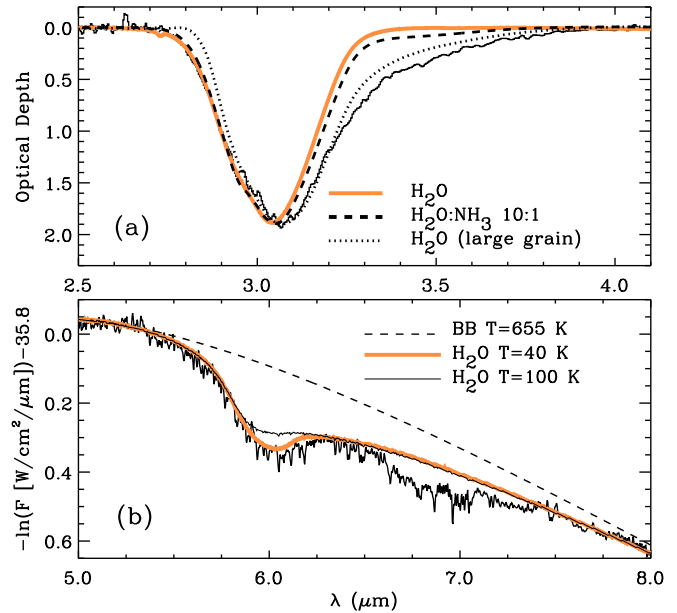,width=250pt,angle=90}
\end{picture}
\caption{Analysis of the 3.0~${\rm \mu m}$ ({\bf a}) and 6.0~${\rm \mu m}$ ({\bf b})
  absorption bands of ${\rm H_2O}$ ice. In panel {\bf a}, the thick
  gray line is a laboratory spectrum of pure ${\rm H_2O}$ ice at
  $T=$40~K. The dashed line is a spectrum of ${\rm H_2O}$:${\rm NH_3}$
  = 10:1 ($T=$ 50~K). The dotted line gives a calculated band profile
  of a large ice grain (see text). Panel {\bf b} shows the 5--8~${\rm
    \mu m}$ spectrum, with laboratory spectra at $T=$ 40~K (thick gray
  line), and at $T$ = 100~K (thin, solid line), showing that only low
  temperatures provide good fits to \object{Elias~29}. The dashed line
  shows the assumed blackbody continuum. The narrow absorption lines
  in the observed spectrum originate from ${\rm H_2O}$ vapor
  (Sect.~\ref{se29:gas}).}~\label{fe29:h2o3_6umfit}
\end{figure}

The continuum determination is complicated by the large width of all
these bands.  In accordance with other studies (Smith et al.
\cite{smi89}; Schutte et al.  \cite{sch96}; Keane et al.
\cite{kea00}), we used single blackbodies to fit the continuum
locally, directly adjacent to the absorption bands.  For the 6~${\rm
  \mu m}$ band we took into account that laboratory spectra of the
bending mode of ${\rm H_2O}$ ice show a prominent wing on the long
wavelength side, extended up to 8~${\rm \mu m}$ (e.g.  Hudgins et al.
\cite{hud93}, Maldoni et al.  \cite{mal98}). We simultaneously fitted
a blackbody continuum, normalized at 5.1~${\rm \mu m}$, and a
laboratory ice spectrum to the observed flux at 8~${\rm \mu m}$ and
the shape of the 6.0~${\rm \mu m}$ feature
(Fig.~\ref{fe29:h2o3_6umfit}).

The shape of the 6.0~${\rm \mu m}$ ${\rm H_2O}$ bending mode is
particularly sensitive to the ice temperature (e.g. Maldoni et al.
\cite{mal98}).  At higher $T$ in the laboratory, the strength of the
main 6.0~${\rm \mu m}$ component decreases at the expense of more
absorption in the long wavelength wing.  In the spectrum of
\object{Elias~29}, the wing cannot be seen as a separate feature,
since at 8~${\rm \mu m}$ it blends with the very deep silicate band.
However, the observed 6.0~${\rm \mu m}$ band is relatively sharp, and
it can only be fitted with ${\rm H_2O}$ ice at $T<80$~K, with a best
fit at $T=40$~K (Fig.~\ref{fe29:h2o3_6umfit}).  The excellent fit to
the 6.0~${\rm \mu m}$ band in \object{Elias~29} indicates that the
5.83 and 6.2~${\rm \mu m}$ excess absorptions detected toward several
massive protostars (Schutte et al. \cite{sch96}, \cite{sch98}; Keane
et al. \cite{kea00}), are not seen in this source (Sect.~3.2.7).

The observed peak position of the stretching mode of ${\rm H_2O}$ ice
toward \object{Elias~29} is 3.07$\pm$0.01 ${\rm \mu m}$. The short
wavelength wing is well matched with a laboratory ice at $T=$40~K, as
for the bending mode (Fig.~\ref{fe29:h2o3_6umfit}).  The long
wavelength wing, however, is poorly fitted.  It has been realized
since long that light scattering by large ice grains leads to extra
extinction on the long wavelength wing (e.g. L\'eger et al.
\cite{leg83}).  To illustrate this, we calculate the extinction cross
section for spherical silicate grains coated with ice mantles,
applying the code given in Bohren \& Huffman (\cite{boh83}) and the
optical constants of Draine \& Lee (\cite{dra84}) and Hudgins et al.
(\cite{hud93}).  Indeed, grains with a core+mantle radius of
$\sim$0.6~${\rm \mu m}$ provide a much better fit to the long
wavelength wing than small grains do (Fig.~\ref{fe29:h2o3_6umfit}).
This effect is unimportant for the 6.0~${\rm \mu m}$ band since it is
intrinsically weaker, and the grains are smaller compared to the
wavelength.  In a more realistic approach, a distribution of grain
sizes, as well as constraints to other observables such as continuum
extinction, the total grain and ice column densities, and polarization
need to be taken into account.  Although there is a general consensus
that large grains need to be invoked (e.g.  Le\'ger et al.
\cite{leg83}; Pendleton et al. \cite{pen90}; Smith et al.
\cite{smi89}; Martin \& Whittet \cite{mar90}), there is no unified
grain model yet that obeys all the observational constraints (e.g.
Smith et al. \cite{smi93}; Tielens \cite{tie82}).  Alternative
absorbers at the long wavelength wing have been proposed, such as
${\rm H_2O}$.${\rm NH_3}$ bondings.  As illustrated in
Fig.~\ref{fe29:h2o3_6umfit}, this effect is however small at the low
column density ratio of ${\rm NH_3}$/${\rm H_2O}$ $<0.13$ toward
\object{Elias~29} (Sect.~3.2.7). A small contribution is also made by
absorption by hydrocarbons (Sect.~3.2.4).

The peak optical depth of the 3.0~${\rm \mu m}$ band is 1.85$\pm$0.08,
which is in excellent agreement with the study of Tanaka et al.
(\cite{tan90}).  Using an integrated band strength $A=2.0\times
10^{-16}$ cm molecule$^{-1}$, we derive a column density of $N$(${\rm
  H_2O}$)= $3.0 \times 10^{18}$ ${\rm cm^{-2}}$ for the small grain
model, and $3.7\times 10^{18}$ ${\rm cm^{-2}}$ for the large grain
model.  Since at present we can not favor one of these two cases, we
will assume an average value of $N$(${\rm H_2O}$)= (3.4$\pm
0.6$)$\times$10$^{18}$ ${\rm cm^{-2}}$ in this paper.  The error bar
also includes the uncertainty in band strength, which increases with
10\% when the ice is heated from 10 to 100~K (Gerakines et al.
\cite{ger95}). Note that a column density determination from the
6.0~${\rm \mu m}$ bending mode is more uncertain due to the unreliable
continuum on the long wavelength side (Fig.~\ref{fe29:h2o3_6umfit}).
At this column density of ${\rm H_2O}$ ice, the depth of the other
vibrational modes is in good agreement with the observed spectrum of
\object{Elias~29} (Fig.~\ref{fe29:waterfit}).

\subsubsection{CO ice} 

\begin{figure}[t!]
\begin{picture}(205,200)(0,0)
\psfig{figure=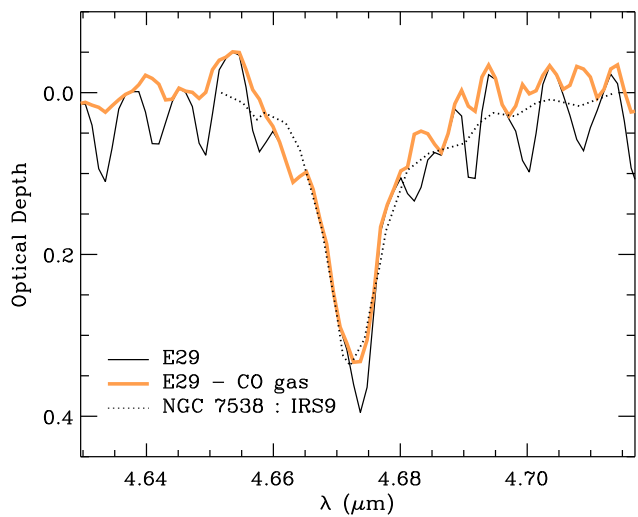,width=240pt,angle=90}
\end{picture}
\caption{The CO ice band on optical depth scale
  observed toward \object{Elias~29} (thin solid line). The thick gray
  line is the spectrum with a gas phase CO model subtracted (see
  text). The dotted line represents the spectrum of the high mass
  protostar \object{NGC~7538~:~IRS9} (divided by 7.5; Tielens et al.
  \cite{tie91}) showing the similarity of the band
  profiles.}~\label{fe29:coice}
\end{figure}

The CO ice band at 4.67~${\rm \mu m}$ in \object{Elias~29} is
contaminated by gas phase CO lines from low $J$ levels
(Figs.~\ref{fe29:coice} and~\ref{fe29:cohighres}). In particular, the
P(1) line lies in the center of the ice band at 4.674~${\rm \mu m}$.
To study the band profile, we subtracted a model for the gaseous lines
at $T_{\rm ex}=750$~K, $N = 5\times10^{18}$~${\rm cm^{-2}}$, and
$b_{\rm D}$=5~${\rm km~s^{-1}}$ (Sect.~\ref{se29:gas}). This increases
the ice band width by $0.9$~${\rm cm^{-1}}$, to FWHM=4.40~${\rm
  cm^{-1}}$ (0.010~${\rm \mu m}$).  With a peak position of
4.673~${\rm \mu m}$ (2140.1~${\rm cm^{-1}}$), the CO ice band observed
toward \object{Elias~29} is similar to that of the luminous protostar
\object{NGC~7538~:~IRS9} (Fig.~\ref{fe29:coice}; Tielens et al.
\cite{tie91}; Chiar et al.  \cite{chi98}). The main, narrow component
at 4.673~${\rm \mu m}$ is attributed to pure solid CO, or CO embedded
in an environment of apolar molecules.  In particular, mixtures with
O$_2$, at an O$_2$/CO ratio as much as 5 (Elsila et al. \cite{els97};
Chiar et al.  \cite{chi98}) provide good fits. Mixtures of CO with
${\rm CO_2}$ are generally too broad (Ehrenfreund et al.
\cite{ehr97}). While the apolar, volatile component dominates the
spectrum, both \object{Elias~29} and \object{NGC~7538~:~IRS9} show
evidence for a wing on the long wavelength side. This is attributed to
CO diluted in a mixture of polar molecules such as ${\rm H_2O}$ and
${\rm CH_3OH}$ (Chiar et al.  \cite{chi98}; Tielens et al.
\cite{tie91}).  Assuming a band strength
$A=1.1\times10^{17}$~cm~molecule~$^{-1}$ for both the polar and apolar
components (Gerakines et al. \cite{ger95}), we derive $N$(CO
ice)=1.7$\times10^{17}$~${\rm cm^{-2}}$ with an apolar/polar ratio of
$\sim$8, comparable to \object{NGC~7538~:~IRS9}. These results are in
good agreement with the ground-based study of Kerr et al.
(\cite{ker93}).  Although \object{NGC~7538~:~IRS9} seems to have a
larger polar CO component in Fig.~\ref{fe29:coice}, this difference
may merely reflect uncertainties in the continuum subtraction, and the
fact that the \object{NGC~7538~:~IRS9} spectrum is not corrected for
gas phase CO lines.

\subsubsection{$\rm CO_2$ ice} 

The absorption bands of ${\rm CO_2}$ ice are prominently present in
the infrared spectrum of \object{Elias~29} (Fig.~\ref{fe29:mir}). We
see the stretching and bending modes at 4.27 and 15.2~${\rm \mu m}$
respectively.  Not visible in this spectrum is the stretching mode of
solid ${\rm ^{13}CO_2}$ at 4.38~${\rm \mu m}$, although the high
resolution spectrum (Fig.~\ref{fe29:cohighres}) shows a hint of its
presence.  A very sensitive observation is presented elsewhere
(Boogert et al. \cite{boo00}).  The ${\rm ^{12}CO_2}$ bending mode and
the ${\rm ^{13}CO_2}$ stretching mode have proven to be very sensitive
to ice mantle composition and thermal history.  In \object{Elias~29},
these bands do not show the narrow substructures seen in many other
protostars, and attributed to heated polar ${\rm CO_2}$ ices (Boogert
et al.  \cite{boo00}; Gerakines et al. \cite{ger99}). As for the CO
ice band (Fig.~\ref{fe29:coice}), the width and peak position of the
${\rm ^{13}CO_2}$ band very much resemble that of the luminous
protostar \object{NGC~7538~:~IRS9}. Thus, the ${\rm CO_2}$ ice toward
\object{Elias~29} is mixed in with polar molecules, and is not much
affected by heating. The ${\rm ^{12}CO_2}$ column density is
22$\pm$4\% relative to ${\rm H_2O}$ ice, which is comparable to the
values reported for high mass protostars (Gerakines et al.
\cite{ger99}).  Finally, we derive an isotope ratio of ${\rm
  ^{12}CO_2}$/${\rm ^{13}CO_2}$=81$\pm$11 in the ice toward
\object{Elias~29}, which is well within the range found for the local
ISM (Boogert et al.  \cite{boo00}).

\subsubsection{The 3.47~${\rm \mu m}$ band} 

\begin{figure}[t!]  
\begin{picture}(250,260)(0,0) 
\psfig{figure=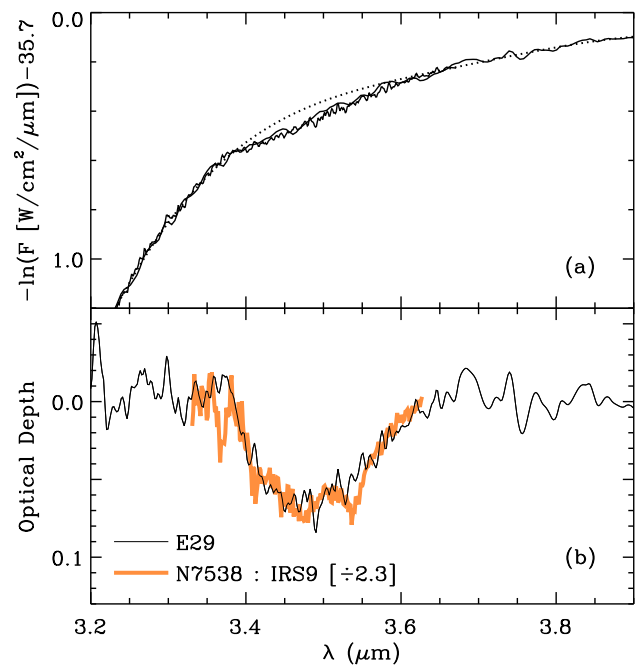,width=250pt,angle=90}
\end{picture} 
\caption{{\bf a and b.} Spectral structure in the long wavelength wing
  of the 3.0~${\rm \mu m}$ band. {\bf a} the merged high ($R=1500$)
  and low ($R=400$) resolution spectra and the assumed polynomial
  continuum (dotted line). {\bf b} optical depth plot of the detected
  3.47~${\rm \mu m}$ feature. The gray, thick line represents the
  ground based spectrum of the high mass protostar
  \object{NGC~7538~:~IRS9}, divided by a factor of 2.3 (Brooke et al.
  \cite{bro99}), showing the C--H stretch mode of solid ${\rm CH_3OH}$
  at 3.54~${\rm \mu m}$.  This feature is absent in the spectrum of
  \object{Elias~29}.}~\label{fe29:347}
\end{figure}

The long wavelength wing of the deep 3.0~${\rm \mu m}$ absorption band
shows a change of slope at 3.38~${\rm \mu m}$, indicative of a shallow
absorption feature (Fig.~\ref{fe29:347}). This feature is also
detected in an independent ground based study of \object{Elias~29}
(Brooke et al. \cite{bro99}).  For consistency with ground based
studies, the continuum on each side of the feature was assumed to
start at 3.37 and 3.61~${\rm \mu m}$.  It must be emphasized however,
that in particular on the long wavelength side, the continuum is
poorly defined.  Fitting a smooth 6-th order polynomial results in an
absorption band centered on 3.49$\pm$0.03~${\rm \mu m}$ with a peak
optical depth of $\tau$=0.06 (Fig.~\ref{fe29:347}). The width is
FWHM=120$\pm$40~${\rm cm^{-1}}$, where the uncertainty includes the
poorly constrained continuum on the long wavelength side.  Features of
similar width and peak position have been detected in several massive
protostellar objects (Allamandola et al. \cite{all92}) and in low mass
objects and quiescent molecular cloud material (Chiar et al.
\cite{chi96}). A likely candidate for this 3.47~${\rm \mu m}$ band is
the C--H stretching mode of hydrocarbons. From the correlation of peak
optical depths of this feature and the 3.0~${\rm \mu m}$ ice band, it
is concluded that the carrier for the 3.47~${\rm \mu m}$ band resides
in ices rather than in refractory dust (Brooke et al. \cite{bro96}).
We find that with $\tau$(3.47~${\rm \mu m}$)=0.06 and $\tau$(3.0~${\rm
  \mu m}$)=1.85, \object{Elias~29} follows this correlation very well.

\subsubsection{CH$_3$OH ice} 

In several high mass protostars, the 3.47~${\rm \mu m}$ band is
blended with a distinct narrow feature centered on 3.54~${\rm \mu m}$
(Allamandola et al. \cite{all92}).  This feature is ascribed to the
C--H stretching mode of solid ${\rm CH_3OH}$. A direct comparison with
the high mass protostar \object{NGC~7538~:~IRS9} shows that, although
the 3.47~${\rm \mu m}$ bands have similar shapes, the 3.54~${\rm \mu
  m}$ feature is absent in \object{Elias~29} (Fig.~\ref{fe29:347}).
We determine a 3$\rm \sigma$ upper limit to the peak optical depth of
$\tau$(3.54~${\rm \mu m}$)$<$0.036. Scaling with the observed depth
and column density in \object{NGC~7538~:~IRS9} (Brooke et al.
\cite{bro99}), then results in an upper limit to the ${\rm CH_3OH}$
ice column density $N$(${\rm CH_3OH}$ ice)$< 1.5\times 10^{17}$ ${\rm
  cm^{-2}}$, or less than 5\% of ${\rm H_2O}$ ice toward
\object{Elias~29} (Table~\ref{te29:abun}). The other modes of ${\rm
  CH_3OH}$ ice are either much weaker, or are severely blended with
the strong ${\rm H_2O}$ and silicate bands (e.g. the C--O stretching
mode at 9.7~${\rm \mu m}$; Schutte et al. \cite{sch91}; Skinner et al.
\cite{ski92}) and thus do not provide better constraints on the ${\rm
  CH_3OH}$ ice column density. Toward other low mass objects, and
quiescent dark clouds, low upper limits have been set to the ${\rm
  CH_3OH}$ ice abundance as well. The ${\rm CH_3OH}$ ice abundance
found in massive protostars is generally of the same magnitude (Chiar
et al. \cite{chi96}), but in a few objects significantly larger
(Dartois et al. \cite{dar99}), than these upper limits.

\begin{figure}[t!]
\begin{picture}(220,203)(0,0)
\psfig{figure=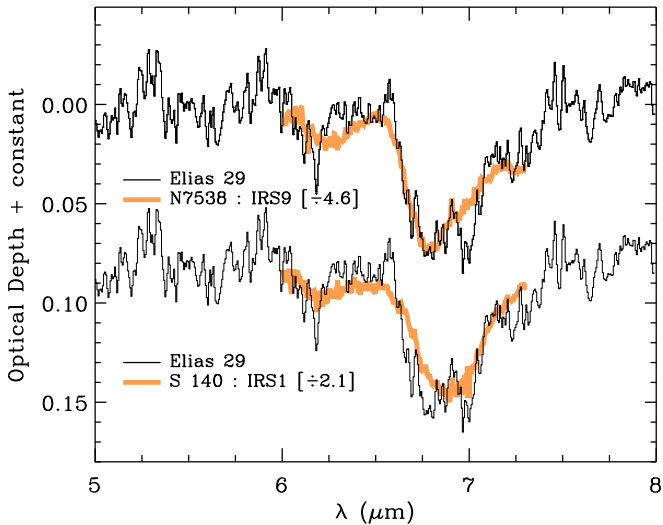,width=250pt,angle=90}
\end{picture}
\caption{Optical depth plot of the 5--8~${\rm \mu m}$ region of \object{Elias~29},
  after subtraction of an ${\rm H_2O}$ ice spectrum at $T=$40~K as
  well as an ${\rm H_2O}$ gas model at $T_{\rm ex}=300$~K,
  $N=2\times10^{18}$~${\rm cm^{-2}}$, $b_{\rm D}$=2.5~${\rm
    km~s^{-1}}$. This figure highlights the 6.85~${\rm \mu m}$
  absorption feature. The thick gray lines give a comparison with the
  massive protostars \object{NGC~7538~:~IRS9} (top) and S~140~:~IRS1
  (bottom; shifted down 0.08 along the optical depth
  axis).}~\label{fe29:68um}
\end{figure}

\subsubsection{The 6.85~${\rm \mu m}$ band}

\object{Elias~29} is the first low mass protostar in which the
6.85~${\rm \mu m}$ absorption band is detected
(Fig.~\ref{fe29:h2o3_6umfit}). After subtraction of the ${\rm H_2O}$
ice band and the gas phase ${\rm H_2O}$ lines (Fig.~\ref{fe29:68um}),
we find that it has a peak optical depth of $\tau \sim$0.07 and an
integrated optical depth $\tau _{\rm int}=7.8\pm 1.6$~${\rm cm^{-1}}$.
When scaled to the ${\rm H_2O}$ ice column density, the strength of
the 6.85~${\rm \mu m}$ band toward \object{Elias~29} is similar to
high mass protostars (Keane et al. \cite{kea00}). The band profile,
e.g.  the sharp edge at 6.60~${\rm \mu m}$, agrees very well with
several high mass objects, in particular those tracing 'cold' gas and
dust (\object{NGC~7538~:~IRS9}, W~33A, GL~989). It clearly deviates
from warmer lines of sight (e.g.  S~140~:~IRS1; Fig.~\ref{fe29:68um}).
Thus, in this picture, we find that the material responsible for the
6.85~${\rm \mu m}$ band toward \object{Elias~29} is not significantly
thermally processed.  Given the low upper limits to the ${\rm CH_3OH}$
ice column density toward \object{Elias~29}, only a fraction of the
band, as for high mass objects, can be explained by the C--H bending
mode of ${\rm CH_3OH}$ ices (Schutte et al. \cite{sch96}). For a
detailed band profile analysis and a discussion on the origin of the
6.85~${\rm \mu m}$ band, we refer to Keane et al. (\cite{kea00}).

\subsubsection{Upper limits to solid $\rm CH_4$, $\rm NH_3$, $\rm
  H_2CO$, HCOOH, OCS, and `XCN'}

Several solid state species have been detected toward luminous
protostars, but are absent toward \object{Elias~29}. The deformation
mode of solid ${\rm CH_4}$ was detected toward protostars, with a peak
position at 1303~${\rm cm^{-1}}$ (7.67~${\rm \mu m}$), and a width
FWHM=11~${\rm cm^{-1}}$ (Boogert et al. \cite{boo96}; Dartois et al.
\cite{dar98b}). For \object{Elias~29} we can exclude this band to a
peak optical depth of $\tau <$0.03, corresponding to $N$(${\rm
  CH_4}$)/$N$(${\rm H_2O}$)$<$1.5\%. This 3$\sigma$ upper limit is
comparable to the detection in \object{NGC~7538~:~IRS9} (Boogert et
al. \cite{boo96}).

\begin{figure}[t!]
\begin{picture}(220,194)(0,0)
\psfig{figure=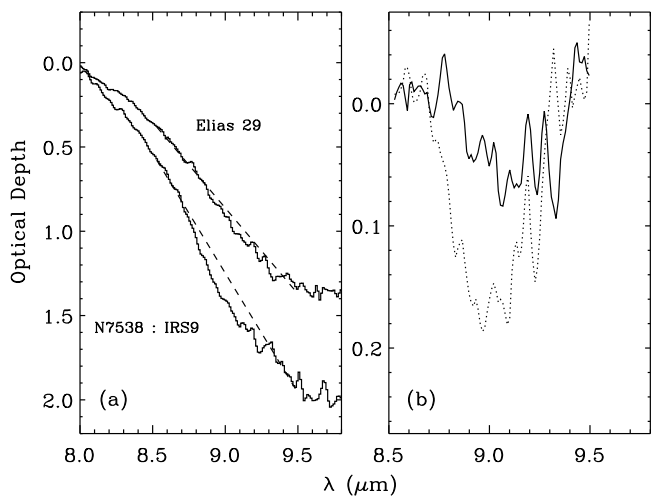,width=250pt,angle=90}
\end{picture}
\caption{{\bf a} ISO--SWS spectra of the silicate band region of
  \object{Elias~29} and \object{NGC~7538~:~IRS9}. The dashed line is
  the local continuum for the solid ${\rm NH_3}$ inversion mode,
  similar to that defined in Lacy et al. (\cite{lac98}).  {\bf b} the
  residuals after continuum subtraction for \object{Elias~29} (solid)
  and \object{NGC~7538~:~IRS9} (dotted). }~\label{fe29:nh3}
\end{figure}

Solid ${\rm NH_3}$ was detected by its 9.10~${\rm \mu m}$ inversion
mode toward \object{NGC~7538~:~IRS9} (Lacy et al. \cite{lac98}). Using
the band strength determined in Kerkhof et al. (\cite{ker99}), the
${\rm NH_3}$ column density is 13\% of ${\rm H_2O}$ ice.  Recent
detections in other highly obscured lines of sight give similar
(W~33A; Gibb et al. \cite{gib00}), or a factor 2 larger ${\rm NH_3}$
abundances (Galactic Center; Chiar et al. \cite{chi00}). To find this
band in the deep silicate feature of \object{Elias~29} we take the
same approach as Lacy et al., by fitting a local straight line
continuum to the wavelength regions 8.52--8.69 and 9.20--9.55~${\rm
  \mu m}$. As a check, we perform the same procedure to the ISO--SWS
spectrum of \object{NGC~7538~:~IRS9} (Fig.~\ref{fe29:nh3}). We confirm
the detection of Lacy et al., although the peak optical depth $\tau
\sim 0.16$ is a factor 2 lower in our case. We ascribe this difference
to the calibration uncertainties of ISO--SWS at this wavelength (Leech
\cite{lee00}). For \object{Elias~29}, a feature with $\tau \sim 0.06$
might be present. However, due to the poorly defined long wavelength
side of the continuum (Fig.~\ref{fe29:nh3}) and the ISO--SWS
calibration uncertainties, we will assume a conservative upper limit
to this band of $\tau<$0.1.  This corresponds to a column density of
$N$(${\rm NH_3}$)$ < 3.5\times10^{17}$${\rm cm^{-2}}$, i.e. $N$(${\rm
  NH_3}$)/$N$(${\rm H_2O}$)$<$13\%.  Other vibrational bands of ${\rm
  NH_3}$ do not provide better constraints.  The equally strong N--H
stretching mode at 2.90~${\rm \mu m}$ (d'Hendecourt \& Allamandola
\cite{hen86}) is hidden in the steep wing of the 3.0~${\rm \mu m}$ ice
band, and there is no significant difference in this region between
the laboratory spectra of pure ${\rm H_2O}$ ice and the mixture ${\rm
  H_2O}$:${\rm NH_3}$=10:1 (Fig.~\ref{fe29:h2o3_6umfit}).  A similar
problem exists for the N--H deformation mode at 6.16~${\rm \mu m}$,
which is hidden in the long wavelength wing of the ${\rm H_2O}$
bending mode (Keane et al. \cite{kea00}). A feature with a peak
optical depth of $\tau <$0.025 would be expected here (Sandford \&
Allamandola \cite{san93}). When subtracting water ice and vapor
absorption, a weak band with an optical depth $\tau =$ 0.03 perhaps
remains present at the expected wavelength (Fig.~\ref{fe29:68um}).
Given the other positive and negative structure in the spectrum, we
regard this also as an upper limit, however.

The ${\rm H_2O}$-subtracted 5.0-6.5~${\rm \mu m}$ wavelength region
(Fig.~\ref{fe29:68um}) does not show the features detected toward high
mass protostars (Schutte et al. \cite{sch96}; \cite{sch98}; Keane et
al. \cite{kea00}). At 6.25~${\rm \mu m}$ (not to confuse with the
feature of ${\rm NH_3}$ ice at slightly shorter wavelength; see
above), a feature has been associated with absorption by carbonaceous
dust (PAH). At 5.83~${\rm \mu m}$ a broad feature has been assigned to
the C=O stretching mode of solid HCOOH, and a narrow feature of solid
H$_2$CO (Keane et al. \cite{kea00}).  Scaling the features observed
toward \object{NGC~7538~:~IRS9} to the lower ${\rm H_2O}$ ice band
column density toward \object{Elias~29}, one would expect peak optical
depths $\tau_{5.83}=0.06$ and $\tau_{6.25}=0.03$.  Our spectra
indicate upper limits to these features of $\tau <$0.03
(Fig.~\ref{fe29:68um}).  Thus, in particular the 5.83~${\rm \mu m}$
feature toward \object{Elias~29} is significantly less pronounced
compared to high mass protostars.  Using the band strengths and
typical widths given in Keane et al. (\cite{kea00}), we derive
3$\sigma$ column density upper limits of $N({\rm H_2CO}) < 6\times
10^{16}$ ${\rm cm^{-2}}$, and $N({\rm HCOOH}) < 3\times 10^{16}$ ${\rm
  cm^{-2}}$. With abundance upper limits of 1--2\% with respect to
${\rm H_2O}$, these aldehydes are thus minor ice components. For
comparison, toward high mass objects it is typically 3\% or higher.

An absorption feature has been detected at 2042$\pm$4~${\rm cm^{-1}}$
(4.90~${\rm \mu m}$) in lines of sight toward several massive
protostars (Palumbo et al. \cite{pal97}).  With a width
FWHM=23$\pm$6~${\rm cm^{-1}}$, it has been ascribed to absorption by
solid OCS. For \object{Elias~29} this feature is not detected with a
peak optical depth $\tau<$0.01 (3$\sigma$), corresponding to
$N$(OCS)$<1.5\times10^{15}$~${\rm cm^{-2}}$ or $<0.05$\% of ${\rm
  H_2O}$ ice. This upper limit is of the same order of magnitude as
the detections in W~33A and Mon~R2~:~IRS2 (Palumbo et al.
\cite{pal97}).

Finally, toward several high and low mass protostars a feature has
been detected at $\sim$2166 ${\rm cm^{-1}}$ (4.62~${\rm \mu m}$) with
a width FWHM$\sim$20~${\rm cm^{-1}}$ (Lacy et al. \cite{lac84}; Tegler
et al. \cite{teg95}). This feature is absent in \object{Elias~29},
with a peak optical depth $\tau<$0.01 (3$\sigma$). If this feature is
caused by the C$\equiv$N stretching mode in `XCN', this corresponds to
a column density $N$(XCN)$<6.7\times10^{15}$~${\rm cm^{-2}}$, or less
than 0.2\% of ${\rm H_2O}$ ice (applying $A=3\times10^{-17}$ cm
molecule$^{-1}$; Tegler et al. \cite{teg95}). This is considerably
less than the detections made toward high mass objects (e.g. W~33A)
and several low mass objects (Elias~18; L~1551~:~IRS5; Tegler et al.
\cite{teg95}). This feature has not been detected in the quiescent
regions of the Taurus molecular cloud (Elias~16;
Table~\ref{te29:abun}). For a more elaborate discussion on this
feature, and the proposed carriers, we refer to Pendleton et al.
(\cite{pen99}).

\subsubsection{Silicates} 

The absorption bands of the Si--O stretching and bending modes of
silicate dust are prominently present at 9.7~${\rm \mu m}$ and
18~${\rm \mu m}$ (Fig.~\ref{fe29:mir}). We derive a peak absorption
optical depth of the 9.7~${\rm \mu m}$ band $\tau _{9.7}=1.38$
(Fig.~\ref{fe29:waterfit}), which is in excellent agreement with the
ground-based study of Hanner et al. (\cite{han95}).  It is likely that
this is a lower limit, since the absorption bands have been partly
filled in with silicate emission from hot dust near the protostar.
Modeling of the 9.7~${\rm \mu m}$ silicate band toward
\object{Elias~29}, including emission and absorption, shows that $\tau
_{9.7}$ ranges between 1.51 and 3.38 for optically thick and thin
emission respectively (Hanner et al.  \cite{han95}). A better fit is
obtained for optically thick emission. In contrast, for luminous
protostars optically thin emission has been generally assumed. Using
the relation $\tau _{9.7}=1.4~\tau _{9.7}{\rm (obs)}+1.6$ (Gillet et
al. \cite{gil75}; Willner et al. \cite{wil82}), yields $\tau
_{9.7}=3.53$ for \object{Elias~29}.

\begin{figure*}[t!]
\begin{picture}(220,215)(-10,0)
\psfig{figure=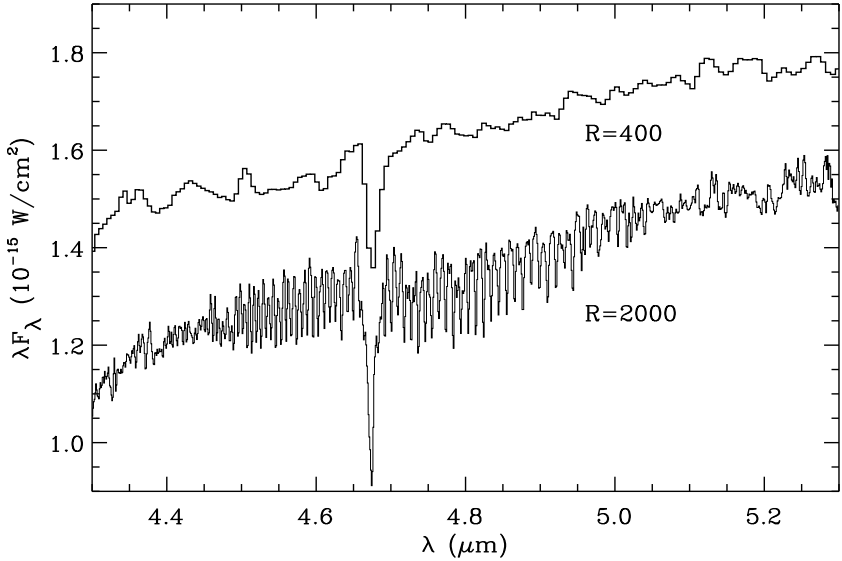,width=320pt,angle=90}
\end{picture}
\hfill \parbox[b]{160pt}{\caption{High resolution ($R$=2000) spectrum
    of \object{Elias~29} showing many gas phase CO lines, and the CO
    ice band at 4.67~${\rm \mu m}$.  The narrow emission line at
    4.653~${\rm \mu m}$ is Pf $\beta$ of \ion{H}{i}. Some of the
    narrow structure seen on the long wavelength side is due to lines
    of hot ${\rm H_2O}$ vapor (see text). For comparison we show a low
    resolution ($R$=400) observation, shifted along the flux scale for
    clarity, where the gas lines are smeared out over the
    continuum.}~\label{fe29:cohighres}}
\end{figure*}

For these values of $\tau _{9.7}$, the visual extinction $A_{\rm V}$
ranges between 28 and 65, assuming the standard relation $A_{\rm
  V}/\tau_{9.7}$=18.5 (Roche \& Aitken \cite{roc84}). However, these
limits are likely overestimated (30-50\%), because of the anomalous
extinction curve due to larger grains in the $\rho$~Oph molecular
cloud (Bohlin et al. \cite{boh78}; Martin \& Whittet \cite{mar90}).
Independent extinction determinations, such as $A_{\rm V}<$48 from the
H--K broad band color and $A_{\rm V}<$80 from C$^{18}$O observations
(Wilking \& Lada \cite{wil83}), do not help to solve this issue.
Millimeter continuum observations (Andr\'e \& Montmerle \cite{and94}),
and the near-infrared J--H color (Greene, priv. comm.), suggest a
relatively low $A_{\rm V}<$30.

The total hydrogen column density $N_{\rm H}=N$(\ion{H}{i})+$2N{\rm
  (H_2)}$ is closely related to $\tau _{9.7}$, and, in contrast to the
derivation of $A_{\rm V}$, the derived $N_{\rm H}$ is not strongly
affected by the large grain size in $\rho$~Oph.  Applying standard
conversion factors for the diffuse ISM (Bohlin et al. \cite{boh78};
Roche \& Aitken \cite{roc84}), we find $N_{\rm
  H}=0.5-1.2\times10^{23}$~${\rm cm^{-2}}$, depending on the applied
$\tau _{9.7}$. To be consistent with studies of high mass protostars,
we will assume in the abundance calculations, the value corresponding
to optically thin silicate emission, i.e. the high limit $N_{\rm
  H}=1.2\times10^{23}$~${\rm cm^{-2}}$ (Table~\ref{te29:abun}).

\subsection{Gas phase absorption lines}~\label{se29:gas}

The high resolution 4.00--8.50~${\rm \mu m}$ spectrum of
\object{Elias~29} shows an impressive number of narrow absorption
lines of gaseous CO and ${\rm H_2O}$ (Figs.~\ref{fe29:h2o3_6umfit}
and~\ref{fe29:cohighres}). We determined local continuum points by
hand and connected these, using a smooth cubic spline interpolation.
Then the data were converted to optical depth scale, and the
absorption lines were modeled, using the ro-vibrational spectra of
gaseous CO and ${\rm H_2O}$ described in Helmich (\cite{hel96}).
These models assume the gas is in Local Thermodynamic Equilibrium
(LTE), and has a single excitation temperature $T_{\rm ex}$. The
absorption lines have a Voigt profile, and are Doppler broadened to a
width $b_{\rm D}$ (=FWHM/$2\sqrt{\rm ln 2}$).  The line oscillator
strengths are calculated from the HITRAN database (Rothman et al.
\cite{rot92}). Finally, the spectrum is convolved with a Gaussian to
the resolution of our observations ($R=1500-2000$).  Thus, three
parameters are varied to fit the observed absorption lines: the column
density $N$, the Doppler parameter $b_{\rm D}$, and the excitation
temperature $T_{\rm ex}$. Reliable column densities can only be
derived if $b_{\rm D}$ is a priori known, which in many studies (like
ours) is not the case, since the lines are unresolved.  At low values
of $b_{\rm D}$, the lines become easily optically thick, and much
larger column densities are needed to fit the observed lines, compared
to models with high $b_{\rm D}$ values, and optically thin lines.

We emphasize that our assumptions of collisional excitation, and LTE
at a single $T_{ex}$ need not be valid.  There is likely a temperature
gradient along the line of sight, as expected for a protostellar
envelope. The LTE assumption may not apply for the high rotational
levels, which have high critical densities. Also, the energy levels
may be pumped by infrared photons, rather than being collisionally
excited. Bearing these caveats in mind, we will here focus on deriving
CO and ${\rm H_2O}$ gas column densities and temperatures using the
LTE models.

\subsubsection{CO gas} 

\begin{figure}[t!]
\begin{picture}(200,195)(-10,0)
\psfig{figure=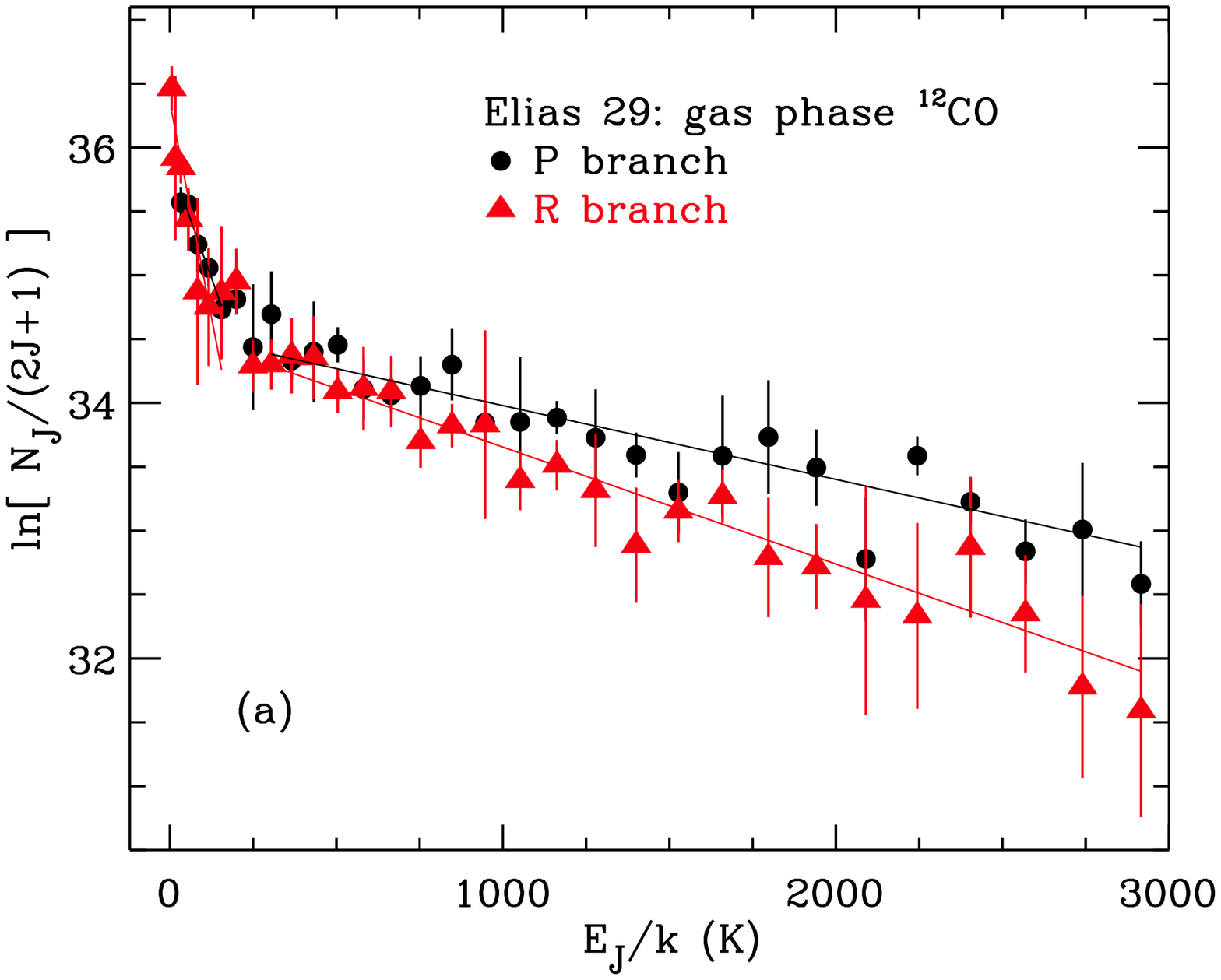,width=240pt,angle=90}
\end{picture}
\begin{picture}(200,195)(0,0)
\psfig{figure=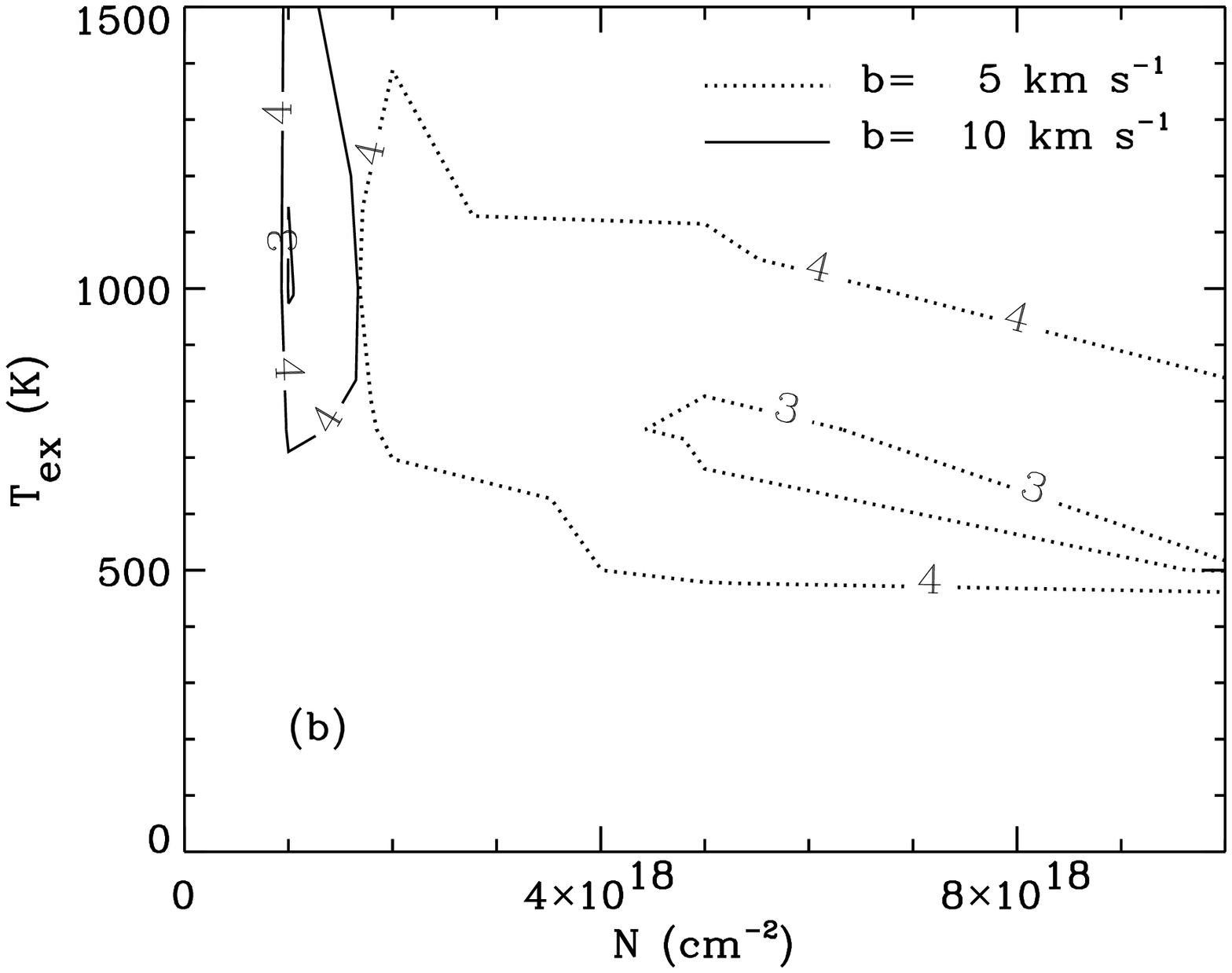,width=250pt,angle=90}
\end{picture}
\caption{{\bf a} rotation diagram of the ${\rm ^{12}CO}$ lines
  detected toward \object{Elias~29} showing the presence of hot and
  cold gas along the line of sight by the different slopes at high and
  low rotational levels. {\bf b} $\chi _{\nu}^2$ contour diagram of
  model fits to the observed ro-vibrational spectrum of the R-branch
  of gaseous CO toward \object{Elias~29}.  $\chi _{\nu}^2$ values are
  shown for the temperature $T_{\rm ex}$ versus CO column density $N$
  at constant velocity broadenings $b_{\rm D}$ of 5~${\rm km~s^{-1}}$
  (dotted) and 10~${\rm km~s^{-1}}$ (solid).  We only show models that
  provide acceptable fits to the data, i.e.  $\chi
  _{\nu}^2<4$.}~\label{fe29:12cochisq}
\end{figure}

\begin{figure}[t!]  
\begin{picture}(200,195)(-10,0) 
\psfig{figure=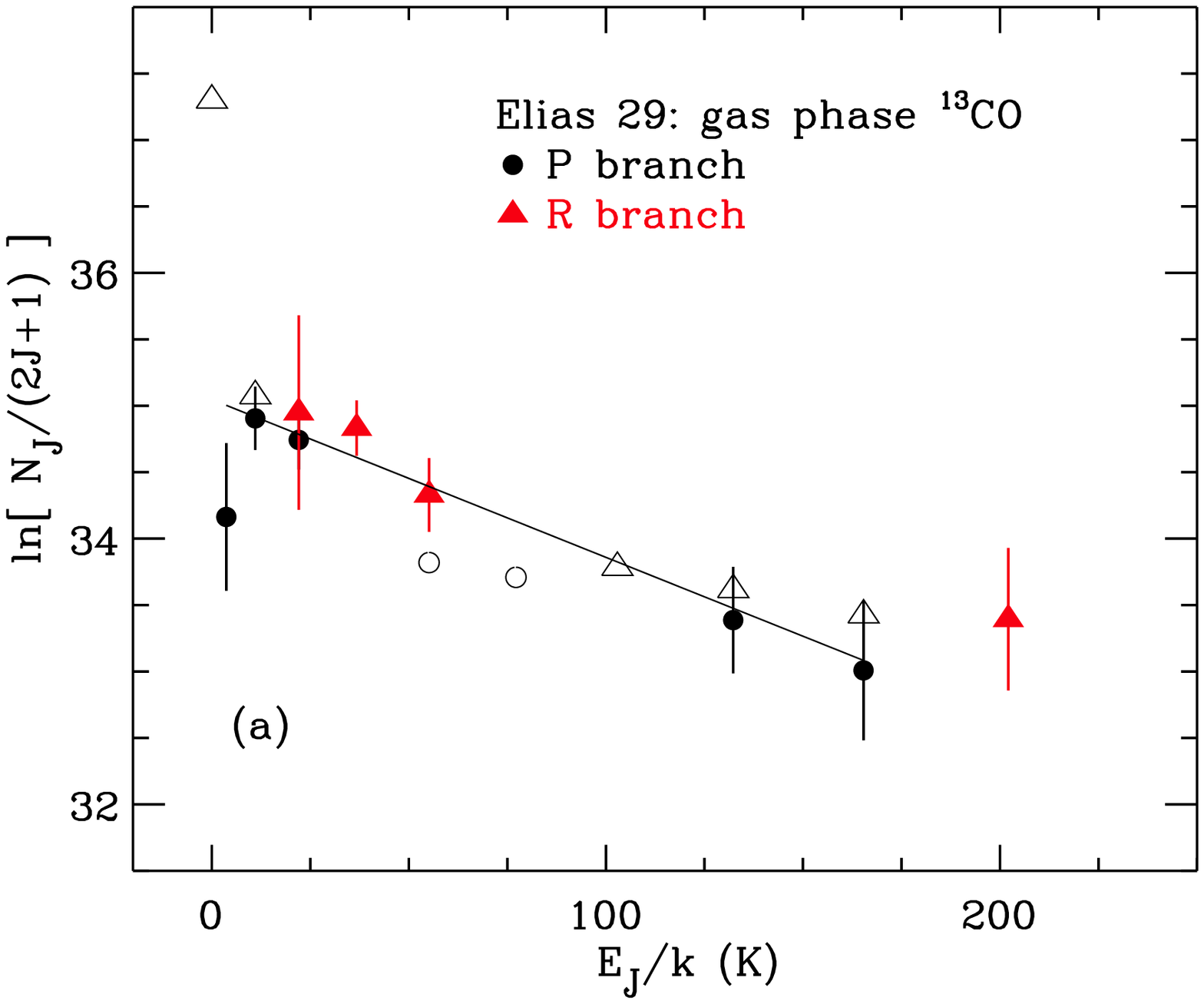,width=240pt,angle=90}
\end{picture} 
\begin{picture}(200,195)(0,0) 
\psfig{figure=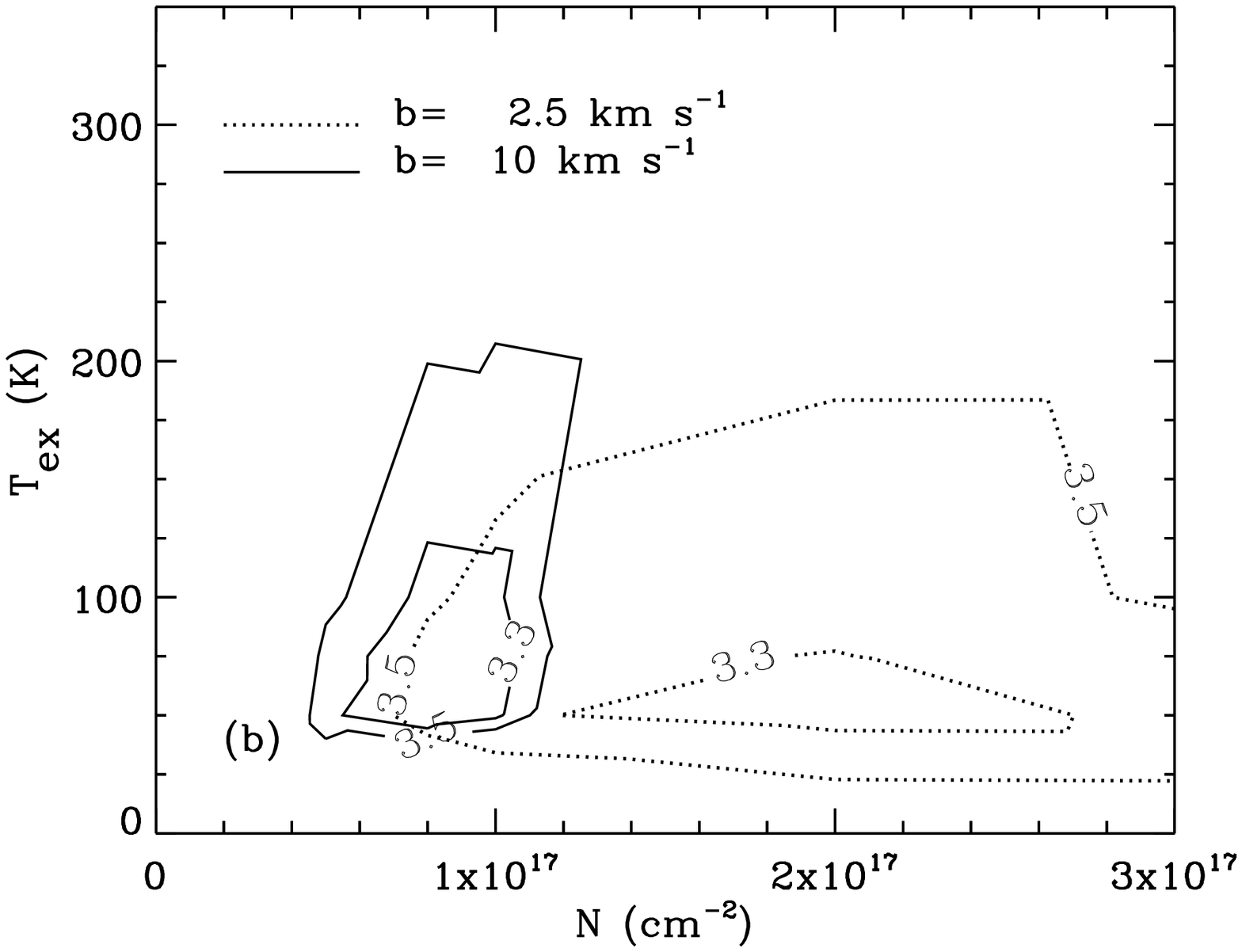,width=250pt,angle=90}
\end{picture} 
\caption{{\bf a} Rotation diagram of the ${\rm ^{13}CO}$ lines
  detected toward \object{Elias~29}. Circles are P-branch lines, and
  triangles are R-branch lines. Open symbols refer to ${\rm ^{13}CO}$
  lines heavily blended with ${\rm ^{12}CO}$ lines, and are not used
  to determine the physical parameters. The straight line indicates
  the best fit, with a gas temperature $T_{\rm rot}=85\pm 57$~K, and a
  column density $N$(${\rm ^{13}CO}$)=$\rm (1.1\pm0.2)\times 10^{17}$
  ${\rm cm^{-2}}$.  {\bf b} $\chi _{\nu}^2$ contour diagram of model
  fits to the observed ro-vibrational spectrum of gaseous ${\rm
    ^{13}CO}$ toward \object{Elias~29}.  $\chi _{\nu}^2$ values are
  shown for the temperature $T_{\rm ex}$ versus CO column density $N$
  for constant velocity broadenings $b_{\rm D}$=2.5~${\rm km~s^{-1}}$
  and $b_{\rm D}$=10~${\rm km~s^{-1}}$. Only acceptable fits to the
  data, having $\chi _{\nu}^2<3.5$, are shown.}~\label{fe29:13cochisq}
\end{figure}

The 4.4--5.0~${\rm \mu m}$ region shows absorption lines of gas phase
${\rm ^{12}CO}$, up to rotational quantum number $J_{\rm low}$=33 in
the R-branch, and $J_{\rm low}$=36 in the P-branch
(Fig.~\ref{fe29:cohighres}). The P(1), P(2) and R(0) lines are blended
with the CO ice band at 4.67~${\rm \mu m}$ and the \ion{H}{i}
Pf~$\beta$ emission line at 4.653~${\rm \mu m}$.  For all other
absorption lines we determined equivalent widths to construct a
rotation diagram.  A rotation diagram gives a first impression of the
temperature components present along the line of sight, as well as
their column densities (or lower limits for optically thick lines).
For technical details on constructing such a diagram we refer to
Mitchell et al.  (\cite{mit90}), and Boogert et al. (\cite{boo98}).
The equivalent widths were converted to column densities, using the
oscillator strengths of Goorvitch (\cite{goo94}). For ${\rm ^{12}CO}$
(Fig.~\ref{fe29:12cochisq}), we find two regimes with very different
slopes, corresponding to temperatures $T_{\rm rot}=90\pm 45$~K and
$T_{\rm rot}=1100\pm 300$~K respectively (with 3$\sigma$ errors).
However, the slopes of the R- and P-branch lines of the hot component
are different (Fig.~\ref{fe29:12cochisq}), resulting in $T_{\rm
  rot}=1700\pm 420$~K when fitting to the P-branch lines only.  A
possible explanation is that CO is excited by continuum photons rather
than collisions. The rising continuum may lead to a higher $T_{\rm
  rot}$ for the P-branch with respect to the R-branch. This effect
becomes stronger when the photons released after de-excitation of
R-branch levels are re-absorbed in P-branch levels. Radiative
excitation has also been used to explain the ${\rm H_2O}$
ro-vibrational spectrum toward Orion BN/KL (Gonzalez-Alfonso et al.
\cite{gon98}). The fact that the ${\rm H_2O}$ P-branch lines are seen
in emission for Orion BN/KL, rather than in absorption as for CO (and
${\rm H_2O}$; Sect. 3.3.2) toward \object{Elias~29}, may reflect a
different density gradient toward \object{Elias~29}, such that the
photons are not able to escape the envelope.  Additionally,
collisional excitation in shocks may be of less importance in
\object{Elias~29} compared to Orion BN/KL. A more careful analysis is
needed to discriminate between the radiative and collisional
excitation models, and alternative explanations, such as non-LTE
effects.

\begin{figure*}[t!]
\begin{picture}(220,235)(-60,0)
\psfig{figure=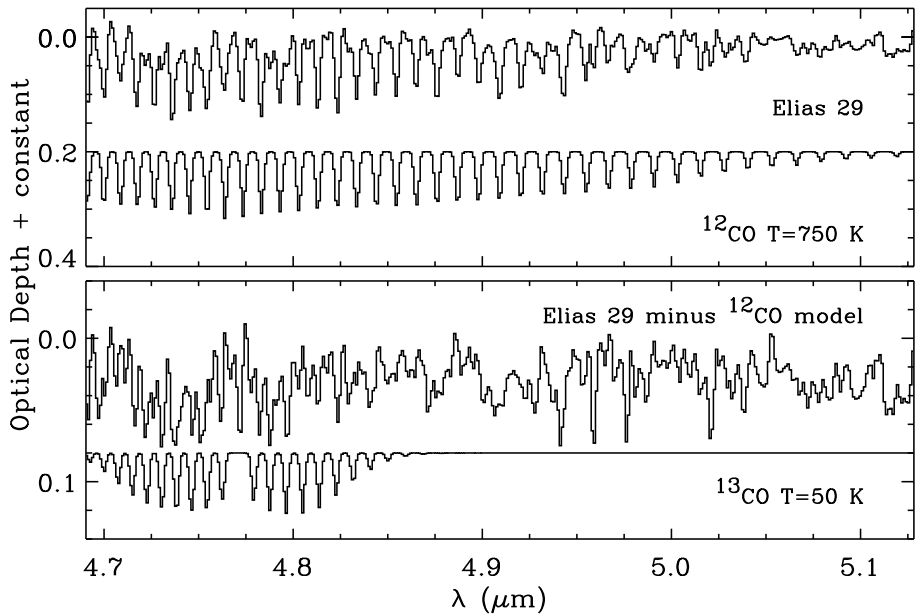,width=350pt,angle=90}
\end{picture}
\caption{The P branch of gas phase CO observed in \object{Elias~29} (top)
  compared with a well fitting ${\rm ^{12}CO}$ gas model at $T_{\rm
    ex}=750$~K ($N=5\times 10^{18}$ ${\rm cm^{-2}}$, $b=5~{\rm
    km~s^{-1}}$).  The bottom panel shows the residual after
  subtraction of the ${\rm ^{12}CO}$ gas model, which contains lines
  of cold ${\rm ^{13}CO}$ gas. For comparison, a ${\rm ^{13}CO}$ model
  is plotted ($T_{\rm ex}=50$~K, $N=2\times 10^{17}$ ${\rm cm^{-2}}$,
  $b=2.5~{\rm km~s^{-1}}$).}~\label{fe29:co_pbranch}
\end{figure*}

The column densities that we derive from the abscissa in the rotation
diagram are $N$(CO)$=1.7\times10^{17}$ and $N$(CO)$=3.5\times10^{17}$
${\rm cm^{-2}}$ for the cold and hot CO components respectively.  To
better constrain the column densities and derive more reliable
temperatures, one has to take into account optical depth effects,
using the LTE model spectra discussed above.  We chose to fit to the
frequency range 2170--2290~${\rm cm^{-1}}$ ($J_{\rm low}> 7$ in
R-branch), thus minimizing the contribution from the cold CO component
and contamination by ${\rm ^{13}CO}$ lines (see below).  We find that
good fits to these high R-branch lines are obtained only for line
widths $b_{\rm D}$$>3$~${\rm km~s^{-1}}$. Sub-millimeter emission line
studies indicate $b_{\rm D}$=3.6~${\rm km~s^{-1}}$ for
CO~J$=6\rightarrow 5$, but much lower values of $b_{\rm D}$=1.2~${\rm
  km~s^{-1}}$ for C$^{18}$O~J$=1\rightarrow 0$ and CS~J$=5\rightarrow
4$ (Boogert, Hogerheijde, et al., in prep.).  Indeed, studies of other
sources have shown that, as a rule, infrared absorption lines are
broader than sub-millimeter emission lines (van der Tak et al.
\cite{tak99}).  Figure~\ref{fe29:12cochisq} shows the $\chi^2_{\nu}$
contour diagram of temperature versus column density for two values of
the line width $b_{\rm D}$=5, and $b_{\rm D}$=10~${\rm km~s^{-1}}$.
The best fitting models have temperatures $T_{\rm ex}=1100\pm$400~K,
in good agreement with the rotation diagram. At $b_{\rm D}$=10~${\rm
  km~s^{-1}}$ the column density is well constrained to
$N$(CO)=(1.3$\pm$ 0.5)$\times 10^{18}$ ${\rm cm^{-2}}$, which is a
factor of 3 larger compared to that derived from the rotation diagram.
Thus at $b_{\rm D}$=10~${\rm km~s^{-1}}$ the lines are still somewhat
optically thick.  At lower $b_{\rm D}$=5~${\rm km~s^{-1}}$, the lines
become very optically thick, and the column density is poorly
constrained.  Although the best fits with $\chi^2_{\nu}$$<3$ have
$N$(CO)=(8$\pm$4)$\times 10^{18}$ ${\rm cm^{-2}}$ at $T_{\rm
  ex}=650\pm 150$~K, reasonable fits are obtained at any
$N$(CO)$>2\times10^{18}$~${\rm cm^{-2}}$ for this hot CO gas.

\begin{table}[t!]
\footnotesize
\begin{center}
\caption{Gas phase ${\rm ^{12}CO}$ and ${\rm H_2O}$ column densities, derived with various 
methods}~\label{te29:cocolden}
\begin{tabular}{llcc}
\noalign{\smallskip} 
\hline 
\noalign{\smallskip} 
Molecule        & Method & \multicolumn{2}{c}{$N$ [$10^{18}$~${\rm cm^{-2}}$]} \\
                &        & cold & hot \\
\noalign{\smallskip} 
\hline
\noalign{\smallskip} 
${\rm ^{12}CO}$ & rotation diagram                          & 0.17            & 0.35 \\ 
${\rm ^{12}CO}$ & LTE, $b_{\rm D}$=10~${\rm km~s^{-1}}$     & --              & 1.3$\pm$0.5 \\ 
${\rm ^{12}CO}$ & LTE, $b_{\rm D}$=5~${\rm km~s^{-1}}$      & --              & $>2$ \\
${\rm ^{13}CO}$ & rotation diagram                          & 9$\pm$2$^{a}$   & -- \\ 
${\rm ^{13}CO}$ & LTE, $b_{\rm D}$=2.5~${\rm km~s^{-1}}$    & 16$\pm$10$^{a}$ & -- \\
\noalign{\smallskip} 
${\rm H_2O}$    & LTE$^{b}$, $b_{\rm D}$=5~${\rm km~s^{-1}}$  & --           & 0.7$\pm$0.4 \\
${\rm H_2O}$    & LTE$^{b}$, $b_{\rm D}$=2.5~${\rm km~s^{-1}}$& --           & 2.4$\pm$2.1 \\
${\rm H_2O}$    & LTE$^{c}$, $b_{\rm D}$=5~${\rm km~s^{-1}}$  & $<1$         & 0.5 \\
${\rm H_2O}$    & LTE$^{c}$, $b_{\rm D}$=2.5~${\rm km~s^{-1}}$& $<10$        & 0.5 \\
\hline
\noalign{\smallskip} 
\multicolumn{4}{p{8cm}}{$^{a}${Converted to $N$(${\rm ^{12}CO}$) assuming
        $N$(${\rm ^{12}CO}$)/$N$(${\rm ^{13}CO}$) = 80 (Boogert et al. \cite{boo00})}}\\
\multicolumn{4}{p{8cm}}{$^{b}${Single temperature model with $T_{\rm
        ex}$=300 K}}\\
\multicolumn{4}{p{8cm}}{$^{c}${Double temperature model with fixed $N_{\rm
        hot}=5\times10^{17}$ ${\rm cm^{-2}}$}}\\
\end{tabular}
\end{center}
\end{table}

\begin{figure*}[t!]  
\begin{picture}(220,235)(-60,0) 
\psfig{figure=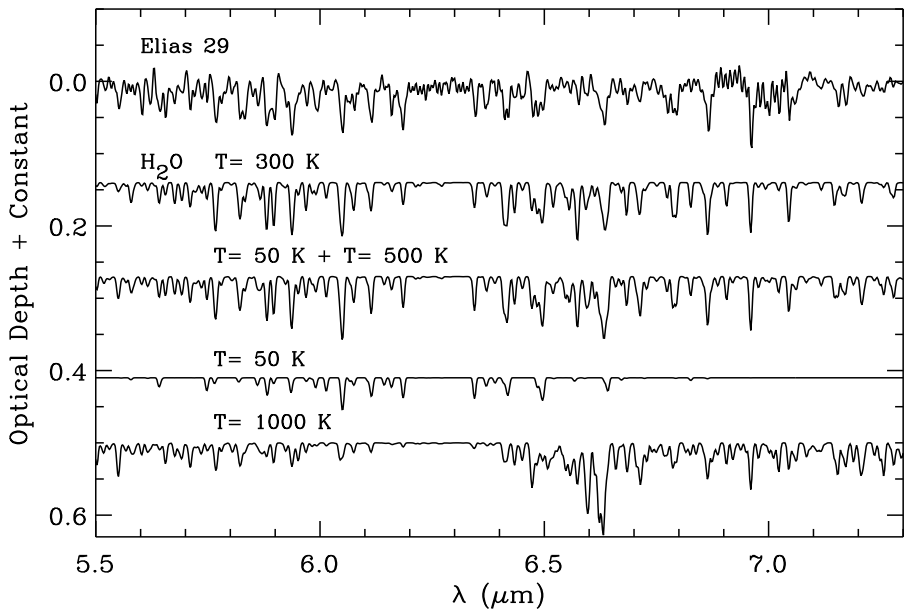,width=350pt,angle=90}
\end{picture} 
\caption{Optical depth plot of gaseous ${\rm H_2O}$ lines observed toward
  \object{Elias~29} (top), compared with model spectra at various
  temperatures. Column densities are $N=5\times10^{17}$ ${\rm
    cm^{-2}}$ for all models, and $b_{\rm D}$=5.0 ${\rm km~s^{-1}}$
  ($T_{\rm ex}=$ 50 K models have $b_{\rm D}$=2.5 ${\rm km~s^{-1}}$).
  The single component $T=300$~K and the two component $T=50+500$~K
  models give equally good fits. The models with just cold ($T=50$~K)
  or very hot ($T=1000$~K) gas clearly do not fit the
  data.}~\label{fe29:h2o_vaporfit}
\end{figure*}

Several ${\rm ^{13}CO}$ lines can be seen in between the ${\rm
  ^{12}CO}$ P-branch lines (Fig.~\ref{fe29:co_pbranch}).  At the
resolution of our observations, the blending with the ${\rm ^{12}CO}$
lines hinders analyzing the much weaker ${\rm ^{13}CO}$ lines.  But
using several well separated lines, we were able to construct a
rotation diagram (Fig.~\ref{fe29:13cochisq}). We find that they result
from cold gas at $T_{\rm rot}=85\pm 57$~K (3$\sigma$ error), in good
agreement with the cold ${\rm ^{12}CO}$ gas temperature. In the
optical thin case, the column density of this cold component is
$N$(${\rm ^{13}CO}$)=$\rm (1.1\pm 0.2)\times 10^{17}$ ${\rm cm^{-2}}$.
However, the detected ${\rm ^{13}CO}$ lines could still be optically
thick.  Therefore, we also modeled the ${\rm ^{13}CO}$ spectrum, and
determine the $\chi^2_{\nu}$ after subtraction of a good fitting hot
${\rm ^{12}CO}$ gas model (Figs.~\ref{fe29:co_pbranch}
and~\ref{fe29:13cochisq}). In the optically thick case, such as for
$b_{\rm D}$=2.5~${\rm km~s^{-1}}$, the column density can have a wide
range $N$(${\rm ^{13}CO}$)=$\rm (2\pm 1.3)\times 10^{17}$ ${\rm
  cm^{-2}}$.  Using the isotope abundance ratio ${\rm ^{12}CO}$/${\rm
  ^{13}CO}$=80 (Boogert et al. \cite{boo00}), the inferred cold ${\rm
  ^{12}CO}$ column density is thus $N$(${\rm ^{12}CO}$)=$(\rm
16\pm10)\times 10^{18}$ ${\rm cm^{-2}}$.  There is also evidence for
${\rm ^{13}CO}$ lines of warm gas ($J_{\rm low}>9$), but at low
significance ($\leq 2\sigma$) and no reliable temperature or column
density could be derived.

We conclude that the CO gas along the line of sight consists of two
temperature components, $T_{\rm rot}=90\pm 45$~K and $T_{\rm
  rot}=1100\pm 300$~K. The column density of both components depends
highly on the assumed line optical thickness
(Table~\ref{te29:cocolden}). Until the intrinsic line width is
directly observed by very high spectral resolution observations, we
can only give a lower limit of $N$(CO--hot)$>2\times10^{18}$ ${\rm
  cm^{-2}}$, while $N$(CO--cold) is not well constrained, i.e. $\rm
(16\pm10)\times 10^{18}$ ${\rm cm^{-2}}$.  Given that $N_{\rm
  H}=1.2\times10^{23}$~${\rm cm^{-2}}$ toward \object{Elias~29}, a
total gas phase CO column density $N$(CO)=$12\times10^{18}$~${\rm
  cm^{-2}}$ is expected, assuming that most of the gas along the line
of sight is molecular and the conversion factor $N(\rm
H_2)/N$(CO)=5000 applies (Lacy et al.  \cite{lac94}).  Then, the ratio
of hot to cold CO gas along the line of sight must be at least 0.2.

\subsubsection{H$_2$O gas} 

We compare the numerous narrow absorption lines detected in the
5-7.3~${\rm \mu m}$ spectral region of \object{Elias~29} with model
spectra of ${\rm H_2O}$ vapor at various physical conditions
(Fig.~\ref{fe29:h2o_vaporfit}).  Clearly, the many lines observed at
wavelengths longer than $\sim$6.55~${\rm \mu m}$ are explained by
${\rm H_2O}$ vapor at a high temperature ($T_{\rm ex}>$100~K). On the
other hand, the relative weakness of the lines observed in the range
6.55--6.65~${\rm \mu m}$ imposes a strict upper limit to the
temperature of this hot gas ($T_{\rm ex}<$1000~K).  To further
constrain the gas temperature, and the ${\rm H_2O}$ vapor column
density, we determined the $\chi^2_{\nu}$ for a large number of
models. Reasonable fits to the full 5--7.3~${\rm \mu m}$ range are
obtained for temperatures of $T_{\rm ex}=350\pm200$ K. The column
density is constrained to $N$=$\rm (7\pm4)\times 10^{17}$ ${\rm
  cm^{-2}}$ for low line optical depths ($b_{\rm D}$$\geq$5). For
narrower lines the column density can be an order of magnitude larger.

\begin{table*}[t!]
\begin{center}
\footnotesize
\caption{Line-of-sight averaged solid and gas phase abundances ($N/N_{\rm H}$ in units of 
$10^{-6}$)}~\label{te29:abun}
\begin{tabular}{lllllll}
\noalign{\smallskip} 
\hline 
\noalign{\smallskip} 
\noalign{\smallskip} 
species         & Dense Cloud$^{a}$   & \object{N7538/9}       & W~33A         & \object{Elias~29}      & GL~2591       & Refs.$^{b}$\\  
\noalign{\smallskip} 
\hline 
\noalign{\smallskip} 
${\rm H_2O}$--ice & 64    & 50    & 39-143        & 28 (8)        & 10            & [1],[2],[3,4],[5],[6]\\ 
\hspace{14pt} --gas$^{c}$
                  & $<$1  & $<3.1$& $<$3.6        & $>$3          & 24 (3)        & [7],[8],[8],[5],[8]\\
CO--ice$^{d}$ [total]
                  & 17 (1)& 8 (1) & 3.2 (1.8)     & 1.4 (0.2)     & $<<1$         & [9],[9],[9],[5],[10]\\ 
\hspace{10pt} --ice 
        [apolar]  & 14 (1)& 7 (0.5)& 0.8 (0.1)    & 1.2 (0.2)     & $<<1$         & [9],[9],[9],[5],[10]\\ 
\hspace{10pt} --gas         
                  & $<10$ & 91 (51)& 143 (32)     &$>67$$^{b}$    &113 (15)       & [11],[12],[12],[5],[12]\\
${\rm CO_2}$--ice & 12 (3)& 10 (1) & 5.2 (0.5)    & 5.4 (0.5)     & 0.9 (0.1)     & [13],[13],[13],[13],[13]\\
\hspace{14pt} --gas       
                  & --    &0.05 (0.01)& 0.08 (0.02)& $<$0.06      & 0.15 (0.03)   & [8], [8], [8], [8]\\
${\rm NH_3}$--ice & --    & 7.6    & 6.1          & $<3.0$        & --            & [14],[3], [5]\\
${\rm CH_3OH}$--ice     
                  & $<1.8$& 2.0    & 7            & $<1.3$        & 4 (2)         & [15],[15],[15],[5],[16]\\
H$_2$CO--ice      & --    & 1.9    & 2.5          & $<0.5$        & --            & [4], [4], [5]\\
HCOOH--ice        & --    & 1.1    & 0.6          & $<0.3$        & --            & [4], [4], [5]\\
${\rm CH_4}$--ice & --    & 0.8    & 0.6          & $<0.4 $       & --            & [17],[17],[5]\\
OCS--ice          & $<0.13$& --    & 0.07         & $<0.02$       & --            & [18],[18],[5]\\
`XCN'--ice        & $<1.3$& 1 (0.3)& 3.6 (1)      & $<0.06$       & --            & [19],[19],[19],[5]\\   
 \noalign{\smallskip} 
\hline 
\noalign{\smallskip} 
$N_{\rm H}~{\rm [10^{23}~cm^{-2}]} ^{e}$ 
                  & 0.39  & 1.6    & 2.8          & 1.2           & 1.7           & [20],[20],[20],[5],[21]\\
\noalign{\smallskip} 
\hline 
\multicolumn{7}{p{15cm}}{$^a$ Ice abundances are for Taurus dense
cloud toward Elias~16. Gas phase abundances are for $\rho$ Oph
cloud.}\\
\multicolumn{7}{p{15cm}}{$^{b}$ References from left to right for each column: 
[1] Chiar et al. \cite{chi95};    
[2] Schutte et al. \cite{sch96};    
[3] Gibb et al. \cite{gib00};  
[4] Keane et al. \cite{kea00};  
[5] this work; 
[6] Smith et al. \cite{smi89};    
[7] Liseau \& Olofsson \cite{lis99};   
[8] Boonman et al. \cite{bood00};   
[9] Chiar et al. \cite{chi98};    
[10] van Dishoeck et al. \cite{dis96};    
[11] Caux et al. \cite{cau99};    
[12] Mitchell et al. \cite{mit90};    
[13] Gerakines et al. \cite{ger99};    
[14] Lacy et al. \cite{lac98};    
[15] Chiar et al. \cite{chi96};    
[16] Schutte et al. \cite{sch91};    
[17] Boogert et al. \cite{boo98};    
[18] Palumbo et al. \cite{pal97};    
[19] Tegler et al. \cite{teg95};    
[20] Tielens et al. \cite{tie91};    
[21] this work, Fig.~\ref{fe29:sedysos}}\\
\multicolumn{7}{p{15cm}}{$^c$ All models assume $b_{\rm D}$=5 ${\rm km~s^{-1}}$ for hot
gas and $b_{\rm D}$=2.5 ${\rm km~s^{-1}}$ for cold gas}\\
\multicolumn{7}{p{15cm}}{$^d$ Total CO ice abundance given as well as
the abundance present in the polar and apolar ice components along the
line of sight}\\
\multicolumn{7}{p{15cm}}{$^e$ Determined from the 9.7~${\rm \mu m}$ silicate
band}\\
\end{tabular}
\end{center}
\end{table*}

In a second approach, we test whether both hot and cold ${\rm H_2O}$
vapor components could be present along the line of sight, much like
the hot and cold CO components. We fitted the regions 5.5--5.8 and
6.55--7.3~${\rm \mu m}$, which do not contain lines from the lowest
rotational levels and thus are particularly sensitive to warm ${\rm
  H_2O}$ vapor along the line of sight (Helmich et al. \cite{held96};
Dartois et al. \cite{dar98b}).  The excitation temperature of this gas
is $T_{\rm ex}=500\pm 300$~K, with column densities similar to that of
the single component model.  In the high temperature regime ($T_{\rm
  ex}\geq$500~K), the modeled line depths in the 6.0--6.5~${\rm \mu
  m}$ region, tracing colder gas, are significantly underestimated.
To determine the temperature and column density of this possible cold
component, we fitted the sum of a good fitting hot gas model ($T_{\rm
  ex}=$500~K, $N=5\times10^{17}$~${\rm cm^{-2}}$, $b_{\rm
  D}$=5.0~${\rm km~s^{-1}}$) and a grid of models at a wide range of
physical conditions to the spectrum of \object{Elias~29}.  Thus, here
we assume that the lines of the hot and cold gas have different radial
velocities and the optical depth spectra can simply be added. We find
that indeed a significant amount of 'cold' ${\rm H_2O}$ vapor, at
$T_{\rm ex}<200$~K may be present (Fig.~\ref{fe29:h2o_vaporfit}).  At
$T_{\rm ex}<100$~K the column density exceeds the assumed hot ${\rm
  H_2O}$ column density of $N=5\times10^{17}$~${\rm cm^{-2}}$.  For a
line width of $b_{\rm D}$=5.0~${\rm km~s^{-1}}$, we find that
$N<1\times10^{18}$~${\rm cm^{-2}}$.  At $b_{\rm D}$=2.5~${\rm
  km~s^{-1}}$, the column density of this cold ${\rm H_2O}$ gas cannot
be constrained.

To summarize, the lines in the 5--7.3~${\rm \mu m}$ range are
reasonably fitted with ${\rm H_2O}$ models at $T_{\rm ex}\sim350\pm
200$~K, and $N$=$\rm (7\pm 4)\times 10^{17}$ ${\rm cm^{-2}}$ at low
line optical depths. For narrower lines ($b_{\rm D}$$<$5~${\rm
  km~s^{-1}}$), the column density can be an order of magnitude
larger. In accordance with the gaseous CO along the line of sight,
equally good fits are obtained with a two component model, where the
cool component has $T_{\rm ex}<200$~K, and the warmer component
$T_{\rm ex}>$500~K. The cool component is then at least as abundant as
the warm ${\rm H_2O}$ gas.

\section{Discussion}~\label{se29:disc} 

\subsection{Gas and solid state abundances}~\label{se29:abun}

We have calculated line of sight averaged gas and solid state
abundances toward \object{Elias~29}, by dividing the column densities
derived in this paper over the total hydrogen column density $N_{\rm
  H}=1.2\times10^{23}$~${\rm cm^{-2}}$
(Sect.~\ref{se29:ice})\footnote{For actual, local, abundances in high
  mass protostars we refer to Boonman et al. \cite{bood00}}. We
compare these abundances with a sample of sight-lines, spanning the
range from dark cloud core to fairly evolved protostars
(Table~\ref{te29:abun}).  As a tracer of ices in quiescent dark cloud
material, we chose the object Elias~16, an evolved star by chance
located behind the Taurus molecular cloud (e.g.  Whittet et al.
\cite{whi98}).  Gas phase CO and ${\rm H_2O}$ abundances in dense
clouds were taken from ISO--LWS studies (Caux et al. \cite{cau99};
Liseau \& Olofsson \cite{lis99}).  The least evolved protostar in our
comparison sample is \object{NGC~7538~:~IRS9}. The infrared spectrum
of this deeply embedded object is characterized by cold ice (Whittet
et al.  \cite{whi96}), and the gas phase temperatures and abundances
indicate a very modest hot core (Mitchell et al.  \cite{mit90};
Boonman et al.  \cite{bood00}). W~33A is more embedded than
\object{NGC~7538~:~IRS9}, but does have a significant amount of warm
gas along the line of sight (Mitchell et al. \cite{mit90}; Lahuis \&
van Dishoeck \cite{lah00}; Boonman et al. \cite{bood00}), and has a
lower abundance of volatile ices (Tielens et al.  \cite{tie91}). The
most evolved object in our sample is GL~2591. It is a typical high
mass hot core source, with low ice abundances and high gas
temperatures. All these protostars are associated with infrared
reflection nebulae, and have well developed high velocity molecular
outflows (Mitchell et al. \cite{mit91}; Bontemps et al. \cite{bon96};
van der Tak et al. \cite{tak00}).  Finally, it is important to note
that all the comparison protostars are at least three orders of
magnitude more luminous than \object{Elias~29}. This allows an
investigation of the effect of low and high mass star formation on the
molecular envelopes.  An extensive comparison with low luminosity
embedded objects is at present not possible, because their infrared
gas and solid state characteristics have not been studied in such
great detail.

The ${\rm H_2O}$ and CO ice abundances decrease for the sequence of
quiescent dense cloud to \object{NGC~7538~:~IRS9}, W~33A and GL~2591
(Table~\ref{te29:abun}).  At the same time, the gas phase ${\rm H_2O}$
abundance, the gas phase CO and ${\rm H_2O}$ temperatures, as well as
the gas-to-solid ratios (Table~\ref{te29:gassol}), increase for these
objects.  All these effects can be explained by evaporation of the ice
mantles and heating of the hot core. It has been suggested that the
observed ${\rm H_2O}$ gas may also have been newly formed by reactions
of atomic O and $\rm H_2$ in warm conditions ($T>200$~K) in the
central hot core or in shocks created by the outflow (e.g., van
Dishoeck \& Blake \cite{disb98}). However, the total (gas plus ice)
${\rm H_2O}$ abundance decreases for the more evolved objects,
indicating that ${\rm H_2O}$ is destroyed rather than being newly
formed (van Dishoeck \cite{dis98}). The low gas phase ${\rm CO_2}$
abundance in all sources indicates that this molecule is destroyed
even more efficiently after evaporation from the grains (Boonman et
al. \cite{bood00}; Charnley \& Kaufman \cite{cha00}).

In the proposed heating sequence, \object{Elias~29} is placed after
W~33A, and before GL~2591.  However, the various ice band profiles
(${\rm H_2O}$, CO, ${\rm CO_2}$, and 6.85~${\rm \mu m}$) in
\object{Elias~29}, indicate little thermal processing, resembling very
much \object{NGC~7538~:~IRS9}, rather than W~33A or GL~2591. The
combination of high gas phase abundances and temperatures, together
with a lack of signatures of thermal processing in the ice bands, as
seen in \object{Elias~29}, is remarkable and is not seen in high mass
protostars.  Geometric effects may play an important role in the
evolution of molecular envelopes around low mass protostars (see
Sect.~\ref{se29:geom}).

Whereas thermal evaporation can explain the abundance variations of
volatiles such as ${\rm H_2O}$, CO, ${\rm CO_2}$, ${\rm NH_3}$, and
${\rm CH_4}$, other mechanisms are needed to explain the variations of
solid ${\rm CH_3OH}$, and XCN abundances among the sources in our
sample (Table~\ref{te29:abun}).  It has been widely considered that
XCN molecules are formed by energetic processing of icy grain mantles
by stellar or cosmic ray induced far-ultraviolet radiation, or by
bombardment with highly energetic particles (e.g. Lacy et al.
\cite{lac84}, Grim \& Greenberg \cite{gri87}, Allamandola et al.
\cite{all88}).  The high ${\rm CH_3OH}$ abundances toward sources with
deep XCN bands, and the apparent absence of ${\rm CH_3OH}$ toward low
mass protostars and dark clouds might suggest that the energetics of
nearby massive stars is needed to produce ${\rm CH_3OH}$ (Gibb et al.
\cite{gib00}).

\begin{table}[h]
\footnotesize
\caption{Gas-to-solid state column density ratios}~\label{te29:gassol}
\begin{tabular}{lllll}
\noalign{\smallskip} 
\hline 
\noalign{\smallskip} 
Object  & CO$^{a}$    & ${\rm H_2O}$$^{a}$  &  ${\rm CO_2}$$^{a}$  & $T_{\rm warm}$$^{b}$\\
\noalign{\smallskip} 
\hline 
\noalign{\smallskip} 
Dark Cl.         & $<$1   & $<$0.02      & --        & --        \\
\object{N7538/9} & 12 (6) & $<$0.05      & 0.005     & 180 (40)  \\
W~33A            & 45 (22)& $<$0.11      & 0.015     & 120 (20)  \\
\object{Elias~29}& $>$53  & $>$0.23      & $<$0.011  & 1000 (500)\\
GL~2591          & $>$400 & 2.4          & 0.17      & 1000 (200)\\
\noalign{\smallskip} 
\hline 
\multicolumn{5}{l}{$^a$ Determined from Table~\ref{te29:abun}}\\
\multicolumn{5}{l}{$^b$ Temperature of warm CO gas (Mitchell et al. \cite{mit90})}\\
\end{tabular}
\end{table}

\subsection{The structure of \object{Elias~29}}~\label{se29:geom}

The variety of dust, gas and ice absorption and emission components
presented here, and in the literature, allows us to construct an
overall view of the structure of \object{Elias~29}. The scale on which
the detected hot CO gas is present can be constrained when one assumes
that the pure rotational high-J CO emission lines detected toward
\object{Elias~29} with ISO--LWS (Ceccarelli et al., in prep.)  are
emitted by the same hot gas.  We fit these observed line fluxes, by
assuming spontaneous, optically thin emission from an LTE level
distribution, and leaving the size of the emitting region as a free
parameter. For the range of column densities and temperatures found to
fit the CO absorption lines (Fig.~\ref{fe29:12cochisq}), we find
diameters in the range 85--225~AU.  Thus, the observed hot CO gas may
be present in a hot core region with the size of a circumstellar disk.
The gas could be concentrated in a high density photospheric layer
above the disk.  To sufficiently heat it by radiation from the central
star, the disk needs to flare outwards, rather than being flat (e.g.
Chiang \& Goldreich \cite{chi97}). We cannot exclude however that the
gas is present more uniformly in the hot core, at lower densities. It
might then also be partly heated by shocks from the outflow close to
the star.  A more detailed modeling of the CO emission lines,
including departures from LTE, optical depth corrections, and taking
into account excitation by radiative pumping, are needed to further
confine the location of the gas phase CO and ${\rm H_2O}$ components
(Ceccarelli et al., in prep.).

The lack of signatures of thermal processing in the ice bands, locates
the ice in a region shielded from the central heating source.  The
ices could be present in a foreground cloud, an extended envelope, or
a circumstellar disk seen close to edge-on.  Millimeter wave continuum
observations indicate an extended envelope, concentrated on the
infrared source (FWHM=17$''$; 2600~AU; Andr\'e \& Montmerle
\cite{and94}; Motte et al. \cite{mot98}).  For a wide range of power
law model fits to the far-infrared SED, Andr\'e \& Montmerle find that
the envelope mass is 0.1~$M_{\odot}$, with a volume averaged dust
temperature of typically $T=$35~K.  This temperature is high enough to
evaporate the most volatile, apolar ices, but too low to induce ice
crystallization. Hence, indeed the observed ${\rm H_2O}$, ${\rm
  CO_2}$, and probably ``6.85''~${\rm \mu m}$ ices could be associated
with this extended envelope. Some of the apolar CO ice has evaporated
in the envelope after the formation of the low mass protostar, as
indicated by the significantly lower CO/${\rm H_2O}$ ice ratio toward
\object{Elias~29}, compared to other sight-lines with little thermal
processing in the ices (\object{NGC~7538~:~IRS9}, Elias~16;
Table~\ref{te29:abun}). In this picture, the detected apolar CO ice
could thus be spatially separate from the other ices, perhaps in
foreground clouds, or well shielded in a very cold disk.

Knowledge of the source structure is essential to interpret the
observed solid and gas phase species. For example, if the ice is
present in the disk, rather than in the envelope, we must see the disk
in a near edge-on configuration.  Is there independent evidence for
the presence of a disk surrounding \object{Elias~29} and what would be
its orientation?  The most direct view is provided by lunar
occultation observations. A central object with diameter of $\sim$1~AU
emits 90\% of the 2.2~${\rm \mu m}$ continuum emission (Simon et al.
\cite{sim87}).  The remaining 10\% comes primarily from an object of
60~AU in diameter, which could be the hot part of a disk ($T\sim
1000$~K).  The strongest spectroscopic disk indicator would be the
presence of emission or absorption of vibrational overtone band heads
of CO (e.g.  Carr \cite{car89}; Najita et al. \cite{naj96}). The
2.0--2.5~${\rm \mu m}$ spectrum of \object{Elias~29} does not show
these features, in contrast to other protostars in $\rho$~Oph, such as
WL~16 (Greene \& Lada \cite{gre96}).  However, the absence of CO
overtone bands does not prove the absence of an (inner) disk (Calvet
et al. \cite{cal91}).  For example, the Herbig Ae object AB~Aur does
not have detected CO overtone bands, while high spatial resolution
radio continuum and emission line observations provide strong evidence
for the presence of a circumstellar disk around this object (Mannings
\& Sargent \cite{man97}).

AB~Aur is an interesting comparison source, since it has the same
luminosity as \object{Elias~29} ($\sim$40~$L_{\odot}$), and the SEDs
of both objects are remarkably similar (Fig.~\ref{fe29:sedysos}).  The
flatness of the SED in AB~Aur is well reproduced in flaring disk
models, where the dust in the outer parts of the disk is more
efficiently heated than in flat disks (Chiang \& Goldreich
\cite{chi97}).  The disk is optically thick up to 100~${\rm \mu m}$,
and becomes optically thin at longer wavelengths where the SED drops
steeply (e.g. van den Ancker et al. \cite{anc00}).  The similarity of
the SEDs does however not necessarily imply that \object{Elias~29} is
dominated by an optically thick disk as well.  A flat SED could also
be produced by the envelope, if it has a shallow power law density
profile (index $\sim 0.5$; Andr\'e \& Montmerle \cite{and94}). This
density profile is remarkably flat compared to high mass protostars
(van der Tak et al. \cite{tak00}; Dartois et al. \cite{dar98a}), and
other low mass protostars (e.g. Hogerheijde \& Sandell \cite{hog00}).
Finally, flat energy distributions are also created by the combination
of a disk and envelope.  Here, the heated envelope irradiates the
outer parts of the disk (Natta \cite{nat93}).

Without direct high resolution imaging, it is difficult to
discriminate between these models. Assuming a given model however, the
present observations put some constraints.  In the disk scenario, its
orientation would have to be closer to edge-on than face-on to explain
the absorption line spectrum of \object{Elias~29} (Chiang \& Goldreich
\cite{chi99}). In these models, an inclination larger than
$\sim$70$^o$ can however be excluded, because this would give an SED
that peaks in the far-infrared, in contrast to what is observed for
\object{Elias~29}.  Also, if the disk were edge-on, a higher absorbing
column, perhaps an order of magnitude larger than the observed $N_{\rm
  H}\sim 1.2\times10^{23}$~${\rm cm^{-2}}$ (Sect.~\ref{se29:ice})
would be expected (Sekimoto et al. \cite{sek97}).  An independent
measure for $N_{\rm H}$ and the disk inclination is provided by the
hard X-ray flux and spectrum, arising from hot gas in the
magnetosphere. For \object{Elias~29}, a high $N_{\rm H}\sim
2\times10^{23}$~${\rm cm^{-2}}$ is observed during X-ray flares, but
$N_{\rm H}$ is a factor of 5 lower in quiescent phases (Kamata et al.
\cite{kam97}). Perhaps the X-ray flares are formed low in the
magnetosphere, and in the relatively high inclination of the disk,
they trace higher column densities compared to X-rays formed in
quiescent phases higher in the magnetosphere.

\section{Conclusions and future work}~\label{se29:summary}

The 1.2--195~${\rm \mu m}$ spectrum of the low mass protostellar
object \object{Elias~29} in the $\rho$~Ophiuchi molecular cloud shows
a wealth of absorption lines of gas and solid state molecules. Hot CO
and ${\rm H_2O}$ gas are detected ($T_{\rm ex}>$300~K) at rather high
abundances, on scales of not more than a few hundred AU. The ice
abundances are relatively low.  In this respect, \object{Elias~29}
resembles luminous protostars with significantly heated cores, such as
GL~2591.  However, {\it none} of the many ice bands that are detected,
i.e. from ${\rm H_2O}$, CO, ${\rm CO_2}$, and the 6.85~${\rm \mu m}$
band, shows outspoken signs of thermal processing. Again in comparison
with luminous protostars, \object{Elias~29} now resembles less evolved
objects, such as \object{NGC~7538~:~IRS9}. Our combined gas and solid
state analysis thus shows that high and low mass protostars heat their
molecular envelopes in different ways.  This may be related to their
different structure, such as the presence of a circumstellar disk in
low mass protostars.  The hot gas of \object{Elias~29} could be
present on the surface of a flaring disk, which is efficiently heated
by the central star.  The ices toward \object{Elias~29} must be well
shielded in a circumstellar disk seen close to edge-on, or far away in
the envelope.

Does this imply that in general the ices in the disks or outer
envelopes of low mass protostars remain unaltered, both in composition
and structure, during the process of star formation? Are these ices
the building blocks of the early solar system and are they preserved
in present day observed cometary nuclei?  To date, no Class~I
protostar has been found with strong signs of crystalline ices
(Boogert et al. \cite{boo00}). On the other hand, the presence of
crystalline ices and silicates has been reported in several isolated,
less embedded Herbig AeBe objects (Malfait et al. \cite{mal99}).  This
research needs to be extended to a larger sample of low mass
protostars, in a range of evolutionary stages and luminosities.
Furthermore, it is essential for the interpretation of the gas and
solid state characteristics toward Elias 29 that the presence of a
circumstellar disk, and its inclination are determined by future high
spatial resolution infrared or millimeter continuum observations.

\begin{acknowledgements} 
  
  We thank Tom Greene (NASA/Ames Research Center) for providing us the
  1.1--2.4~${\rm \mu m}$ spectrum of \object{Elias~29} in electronic
  format, Willem Schutte (Leiden Observatory) for providing the ${\rm
    H_2O}$:${\rm NH_3}$ laboratory ice mixtures, and T.Y. Brooke
  (NASA/JPL) for the 3~${\rm \mu m}$ spectrum of
  \object{NGC~7538~:~IRS9}. The referee D. Ward-Thompson is thanked
  for a number of useful comments. D.C.B.W. is funded by NASA through
  JPL contract no.961624 and by the NASA Exobiology and Long-Term
  Space Astrophysics programs (grants NAG5-7598 and NAG5-7884,
  respectively).

\end{acknowledgements}



\begin{thebibliography}{}

\bibitem[1987]{ada87}Adams F.C., Lada C.J., Shu F.H., 1987, ApJ 312,

\bibitem[1988]{all88}Allamandola L.J., Sandford S.A., Valero G.J.,
1988, Icarus 76, 225

\bibitem[1992]{all92}Allamandola L.J., Sandford S.A., Tielens
A.G.G.M., Herbst T.M., 1992, ApJ 399, 134

\bibitem[1994]{and94}Andr\'e Ph., Montmerle Th., 1994, ApJ 420, 837

\bibitem[1993]{and93}Andr\'e Ph., Ward-Thompson D., Barsony M., 1993,
ApJ 406, 122

\bibitem[1978]{boh78}Bohlin R.C., Savage B.D., Drake J.F., 1978, ApJ
224, 132

\bibitem[1983]{boh83}Bohren C.F., Huffman D.R., 1983, Absorption and
Scattering of Light by Small Particles. John Wiley \& Sons, New York,
App. B

\bibitem[1996]{bon96}Bontemps S., Andr\'e P., Terebey S., Cabrit S.,
1996, A\&A 311, 858

\bibitem[1996]{boo96}Boogert A.C.A, Schutte W.A., Tielens A.G.G.M., et
al., 1996, A\&A 315, L377

\bibitem[1998]{boo98}Boogert A.C.A., Helmich F.P., van Dishoeck E.F.,
et al., 1998, A\&A 336, 352

\bibitem[2000]{boo00}Boogert A.C.A., Ehrenfreund P., Gerakines P.A.,
et al., 2000, A\&A 353, 349

\bibitem[2000]{bood00}Boonman A.M.S., van Dishoeck E.F., Doty S.D.,
et al., 2000, A\&A, subm.

\bibitem[1996]{bro96}Brooke T.Y., Sellgren K., Smith R.G., 1996, ApJ
459, 209

\bibitem[1999]{bro99}Brooke T.Y., Sellgren K., Geballe T.R., 1999, ApJ
517, 883

\bibitem[1991]{cal91}Calvet N., Patino A., Magris G.C., d'Alessio P.,
1991, ApJ 380, 617

\bibitem[1989]{car89}Carr J.S., 1989, ApJ 345, 522

\bibitem[1999]{cau99}Caux E., Ceccarelli C., Castets A., et al., 1999,
A\&A 347, L1

\bibitem[2000]{cha00}Charnley S.B., Kaufman M.J., 2000, ApJ 529, 111

\bibitem[1995]{che95}Chen H., Myers P.C., Ladd E.F., Wood D.O.S.,
1995, ApJ 445, 377

\bibitem[1997]{chi97}Chiang E.I., Goldreich P., 1997, ApJ 490, 368

\bibitem[1999]{chi99}Chiang E.I., Goldreich P., 1999, ApJ 519, 279

\bibitem[1995]{chi95}Chiar J.E., Adamson A.J., Kerr T.H., Whittet
D.C.B., 1995, ApJ 455, 234

\bibitem[1996]{chi96}Chiar J.E., Adamson A.J., Whittet D.C.B., 1996,
ApJ 472, 665

\bibitem[1998]{chi98}Chiar J.E., Gerakines P.A., Whittet D.C.B., et
al., 1998, ApJ 498, 716

\bibitem[2000]{chi00}Chiar J.E., Tielens A.G.G.M., Whittet D.C.B., et
al., 2000, ApJ 537, in press

\bibitem[1996]{cle96}Clegg P.E., Ade P.A.R., Armand C., et al., 1996,
A\&A 315, L38

\bibitem[1998a]{dar98a}Dartois E., Cox P., Roelfsema P.R., et al.,
1998a, A\&A 338, 21

\bibitem[1998b]{dar98b}Dartois E., d'Hendecourt L., Boulanger F., et
al., 1998b, A\&A 331, 651

\bibitem[1999]{dar99}Dartois E., Schutte W., Geballe T.R., et al.,
1999, A\&A 342, 32

\bibitem[1996]{gra96}de Graauw Th., Haser L.N., Beintema D.A., et al.,
1996, A\&A 315, L49

\bibitem[1986]{hen86}d'Hendecourt L.B., Allamandola L.J., 1986, A\&AS
64, 453

\bibitem[1984]{dra84}Draine B.T., Lee H.M., 1984, ApJ 285, 89

\bibitem[1997]{ehr97}Ehrenfreund P., Boogert A.C.A., Gerakines P.A.,
Tielens A.G.G.M., van Dishoeck E.F., 1997, A\&A 328, 649

\bibitem[1978]{eli78}Elias J.H., 1978, ApJ 224, 857

\bibitem[1997]{els97}Elsila J., Allamandola L.J., Sandford S.A., 1997,
ApJ 479, 818

\bibitem[1995]{ger95}Gerakines P.A., Schutte W.A., Greenberg J.M., van
Dishoeck E.F., 1995, A\&A 296, 810

\bibitem[1999]{ger99}Gerakines P.A., Whittet D.C.B., Ehrenfreund P.,
1999, ApJ 522, 357

\bibitem[2000]{gib00}Gibb E.L., Whittet D.C.B., Schutte W.A., et al.,
2000, ApJ, in press

\bibitem[1975]{gil75}Gillett F.C., Forrest W.J., Merrill K.M., Soifer
B.T., Capps R.W., 1975, ApJ 200, 609

\bibitem[1998]{gon98}Gonzalez-Alfonso E., Cernicharo J., van Dishoeck
E.F., Wright C.M., Heras A., 1998, ApJ 502, L169

\bibitem[1994]{goo94}Goorvitch D., 1994, ApJS 95, 535

\bibitem[1996]{gre96}Greene Th.P., Lada C.J., 1996, AJ 112, 2184

\bibitem[2000]{gre00}Greene Th.P., Lada C.J., 2000, AJ, in press

\bibitem[1987]{gri87}Grim R.J.A., Greenberg J.M., 1987, ApJ 321, L91

\bibitem[1995]{han95}Hanner M.S., Brooke T.Y., Tokunaga A.T., 1995,
ApJ 438, 250

\bibitem[1996]{hel96}Helmich F.P., 1996,
Ph. D. thesis. Rijksuniversiteit Leiden

\bibitem[1996]{held96}Helmich F.P., van Dishoeck E.F., Black J.H., et
al., 1996, A\&A 315, L173

\bibitem[1992]{hil92}Hillenbrand L.A., Strom S.E., Vrba F.J., Keene
J., 1992, ApJ 397, 613

\bibitem[2000]{hog00}Hogerheijde M.R., Sandell G., 2000, ApJ, in press

\bibitem[1993]{hud93}Hudgins D.M., Sandford S.A., Allamandola L.J.,
Tielens A.G.G.M., 1993, ApJS 86, 713

\bibitem[1997]{ive97}Ivezic Z., Elitzur M., 1997, MNRAS 287, 799

\bibitem[1997]{kam97}Kamata Y., Koyama K., Tsuboi Y., Yamauchi S.,
1997, PASJ 49, 461

\bibitem[2000]{kea00}Keane J.V., Tielens A.G.G.M., Boogert A.C.A.,
Schutte W.A., Whittet D.C.B., 2000, A\&A, in press

\bibitem[1999]{ker99}Kerkhof O., Schutte W.A., Ehrenfreund P., 1999,
A\&A 346, 990

\bibitem[1993]{ker93}Kerr T.H., Adamson A.J., Whittet D.C.B., 1993,
MNRAS 262, 1047

\bibitem[1996]{kes96}Kessler M.F., Steinz J.A., Anderegg M.E., et al.,
1996, A\&A 315, L27

\bibitem[1984]{lac84}Lacy J.H., Baas F., Allamandola L.J., et al.,
1984, ApJ 276, 533

\bibitem[1994]{lac94}Lacy J.H., Knacke R., Geballe T.R., Tokunaga
A.T., 1994, ApJ 428, L69

\bibitem[1998]{lac98}Lacy J.H., Faraji H., Sandford S.A., Allamandola
L.J., 1998, ApJ 501, 105

\bibitem[1984]{lad84}Lada C.J., Wilking B.A., 1984, ApJ 287, 610

\bibitem[2000]{lah00}Lahuis F., van Dishoeck E.F., 2000, A\&A 355, 699

\bibitem[2000]{lee00}Leech K., 2000, The ISO handbook, vol.VI: SWS,
{http://isowww.estec.esa.nl/manuals/handbook/vi/sws\_hb/}, p.132

\bibitem[1983]{leg83}L\'eger A., Gauthier S., Defourneau D., Rouan D.,
1983, A\&A 117, L164

\bibitem[1999]{lis99}Liseau R., Olofsson G., 1999, A\&A 343, L83

\bibitem[1990]{lor90}Loren R.B., Wootten A., Wilking B.A., 1990, ApJ
365, 269

\bibitem[1998]{mal98}Maldoni M.M., Smith R.G., Robinson G., Rookyard
V.L., 1998, MNRAS 298, 251

\bibitem[1999]{mal99}Malfait K., Waelkens C., Bouwman J., de Koter A.,
Waters L.B.F.M., 1999, A\&A 345, 181

\bibitem[1994]{man94}Mannings V., 1994, MNRAS 271, 587

\bibitem[1997]{man97}Mannings V., Sargent A.I., 1997, ApJ 490, 792

\bibitem[1990]{mar90}Martin P.G., Whittet D.C.B., 1990, ApJ 357, 113

\bibitem[1990]{mit90}Mitchell G.F., Maillard J.-P., Allen M., Beer R.,
Belcourt K., 1990, ApJ 363, 554

\bibitem[1991]{mit91}Mitchell G.F., Maillard J.-P., Hasegawa T.I.,
1991, ApJ 371, 342

\bibitem[1998]{mot98}Motte F., Andr\'e Ph., Neri R., 1998, A\&A 336,
150

\bibitem[1996]{naj96}Najita J., Carr J.S., Tokunaga A.T., 1996, ApJ
456, 292

\bibitem[1993]{nat93}Natta A., 1993, ApJ 412, 761

\bibitem[1993]{pal93}Palla F., Stahler S.W., 1993, ApJ 418, 414

\bibitem[1997]{pal97}Palumbo M.E., Geballe T.R., Tielens A.G.G.M.,
1997, ApJ 479, 839

\bibitem[1990]{pen90}Pendleton Y.J., Tielens A.G.G.M., Werner M.W.,
1990, ApJ 349, 107

\bibitem[1999]{pen99}Pendleton Y.J., Tielens A.G.G.M., Tokunaga A.T.,
Bernstein M.P., 1999, ApJ 513, 294

\bibitem[1984]{roc84}Roche P.F., Aitken D.K., 1984, MNRAS 208, 481

\bibitem[1992]{rot92}Rothman L.S., Gamache R.R., Tipping R.H., et al.,
1992, J. Quant. Spectrosc. Radiat. Transfer 48, 469

\bibitem[1993]{san93}Sandford S.A., Allamandola L.J., 1993, ApJ 417,
815

\bibitem[1996]{sar96}Saraceno P., Andr\'e P., Ceccarelli C., Griffin
M., Molinari S., 1996, A\&A 309, 827

\bibitem[1991]{sch91}Schutte W.A., Tielens A.G.G.M., Sandford S.A.,
1991, ApJ 382, 523

\bibitem[1996]{sch96}Schutte W.A., Tielens A.G.G.M., Whittet D.C.B.,
et al., 1996, A\&A 315, 333

\bibitem[1998]{sch98}Schutte W.A., van der Hucht K.A., Whittet D.C.B.,
et al., 1998, A\&A 337, 261

\bibitem[1997]{sek97}Sekimoto Y., Tatematsu K., Umemoto T. et al.,
1997, ApJ 489, L63

\bibitem[1987]{sim87}Simon M., Howell R.R., Longmore A.J., et al.,
1987, ApJ 320, 344

\bibitem[1992]{ski92}Skinner C.J., Tielens A.G.G.M., Barlow M.J.,
Justtanont K., 1992, ApJ 399, 79

\bibitem[1989]{smi89}Smith R.G., Sellgren K., Tokunaga A.T., 1989, ApJ
344, 413

\bibitem[1993]{smi93}Smith R.G., Sellgren K., Brooke T.Y., 1993, MNRAS
263, 749

\bibitem[1996]{swi96}Swinyard B.M., Clegg P.E., Ade P.A.R., et al.,
1996, A\&A 315, L43

\bibitem[1990]{tan90}Tanaka M., Sato S., Nagata T., Yamamoto T., 1990,
ApJ 352, 724

\bibitem[1995]{teg95}Tegler S.C., Weintraub D.A., Rettig T.W., et al.,
1995, ApJ 439, 279

\bibitem[1982]{tie82}Tielens A.G.G.M., 1982,
Ph. D. thesis. Rijksuniversiteit Leiden

\bibitem[1991]{tie91}Tielens A.G.G.M., Tokunaga A.T., Geballe T.R.,
Baas F. , 1991, ApJ 381, 181

\bibitem[2000]{anc00}van den Ancker M.E., Bouwman J., Wesselius P.R.,
et al., 2000, A\&A, in press

\bibitem[1999]{tak99}van der Tak F.F.S., van Dishoeck E.F., Evans
N.J. II, Bakker E.J., Blake G.A., 1999, ApJ 522, 991

\bibitem[2000]{tak00}van der Tak F.F.S., van Dishoeck E.F., Evans
N.J. II, Blake G.A., 2000, ApJ, in press

\bibitem[1998]{dis98}van Dishoeck E.F., 1998, Faraday Discussions 109,
31

\bibitem[1998]{disb98}van Dishoeck E.F., Blake G.A., 1998, ARAA 36,
317

\bibitem[1996]{dis96}van Dishoeck E.F., Helmich F.P., de Graauw Th.,
et al., 1996, A\&A 315, L349

\bibitem[1974]{whi74}Whittet D.C.B., 1974, MNRAS 168, 371

\bibitem[1996]{whi96}Whittet D.C.B., Schutte W.A., Tielens A.G.G.M.,
et al., 1996, A\&A 315, L357

\bibitem[1998]{whi98}Whittet D.C.B., Gerakines P.A., Tielens A.G.G.M.,
et al., 1998, ApJ 498, L159

\bibitem[1983]{wil83}Wilking B.A., Lada C.J., 1983, ApJ 271, 698

\bibitem[1989]{wil89}Wilking B.A., Lada C.J., Young E.T., 1989, ApJ
340, 823

\bibitem[1982]{wil82}Willner S.P., Gillet F.C., Herter T.L., et al.,
1982, ApJ 253, 174

\bibitem[1986]{you86}Young E.T., Lada C.J., Wilking B.A., 1986, ApJ
304, L45

\end{thebibliography}
\end{document}